\documentclass[%
notitlepage,
superscriptaddress,
twocolumn,
amsmath,amssymb,
aps,
prb,
10pt
]{revtex4-2}

\usepackage[utf8]{inputenc}
\usepackage{graphicx}			
\usepackage{dcolumn}			
\usepackage{bm}					
\usepackage{amssymb}
\usepackage{epsfig}
\usepackage{gensymb}
\usepackage{multirow}
\usepackage{array}
\usepackage{ctable}
\usepackage{hyperref}
\usepackage{textcomp}
\usepackage{gensymb}
\usepackage[immediate]{silence}

\makeatletter
\g@addto@macro\maketitle{\thispagestyle{plain}}  
\makeatother

\begin{document}

\title{Measuring the magnon-photon coupling in shaped ferromagnets: tuning of the resonance frequency}

\author{Sergio Mart\'inez-Losa del Rinc\'on}
\affiliation{Instituto de Nanociencia y Materiales de Arag\'on (INMA), CSIC-Universidad de Zaragoza, Zaragoza, Spain}
\affiliation{Departamento de F\'isica de la Materia Condensada, Universidad de Zaragoza}

\author{Ignacio Gimeno}
\affiliation{Instituto de Nanociencia y Materiales de Arag\'on (INMA), CSIC-Universidad de Zaragoza, Zaragoza, Spain}
\affiliation{Departamento de F\'isica de la Materia Condensada, Universidad de Zaragoza}

\author{Jorge P\'erez-Bail\'on}
\affiliation{Instituto de Nanociencia y Materiales de Arag\'on (INMA), CSIC-Universidad de Zaragoza, Zaragoza, Spain}
\affiliation{Departamento de F\'isica de la Materia Condensada, Universidad de Zaragoza}

\author{Victor Rollano}
\affiliation{Instituto de Nanociencia y Materiales de Arag\'on (INMA), CSIC-Universidad de Zaragoza, Zaragoza, Spain}
\affiliation{Departamento de F\'isica de la Materia Condensada, Universidad de Zaragoza}

\author{Fernando Luis}
\affiliation{Instituto de Nanociencia y Materiales de Arag\'on (INMA), CSIC-Universidad de Zaragoza, Zaragoza, Spain}
\affiliation{Departamento de F\'isica de la Materia Condensada, Universidad de Zaragoza}

\author{David Zueco}
\affiliation{Instituto de Nanociencia y Materiales de Arag\'on (INMA), CSIC-Universidad de Zaragoza, Zaragoza, Spain}
\affiliation{Departamento de F\'isica de la Materia Condensada, Universidad de Zaragoza}

\author{Mar\'ia Jos\'e Mart\'inez-P\'erez}
\email{pemar@unizar.es}
\affiliation{Instituto de Nanociencia y Materiales de Arag\'on (INMA), CSIC-Universidad de Zaragoza, Zaragoza, Spain}
\affiliation{Departamento de F\'isica de la Materia Condensada, Universidad de Zaragoza}


\begin{abstract}
Cavity photons and ferromagnetic spins excitations can exchange information coherently in hybrid architectures, at speeds set by their mutual coupling strength. Speed enhancement is usually achieved by optimizing the geometry of the electromagnetic cavity. Here we show that the geometry of the ferromagnet plays also an important role, by setting the fundamental frequency of the magnonic resonator. Using focused ion beam patterning, we vary the aspect ratio of different Permalloy samples reaching operation frequencies above 10 GHz  while working at low external magnetic fields. Additionally, we perform broad band ferromagnetic resonance measurements and cavity experiments that demonstrate that the light-matter coupling strength can be estimated using either open transmission lines or resonant cavities, yielding very good agreement. Finally, we describe a simple theoretical framework based on electromagnetic and micromagnetic simulations that successfully accounts for the experimental results. This approach can be used to design hybrid quantum systems exploiting whatsoever magnetostatic mode excited in ferromagnets of arbitrary size and shape and to tune their operation conditions.
\end{abstract}



\maketitle

\section{Introduction}

In recent years, quantum manipulation and read-out protocols based on circuit quantum electrodynamics \cite{Blais2007} have been successfully applied to the field of quantum magnonics \cite{Harder2021,Li2020,LachanceQuirion2019}. In these studies, a coherent coupling between photons in an electromagnetic cavity and magnons in a ferromagnet is achieved in the strong \cite{Tabuchi2014,Huebl2013}, ultra-strong \cite{Golovchanskiy2021,Flower2019,Zhang2014} and, even, close to the deep-strong regimes \cite{Golovchanskiy2021a}. 
This has opened the way to the observation of interesting phenomena with strong potential for applications like quantum transduction between optical and microwave photons \cite{Lambert2020,Hisatomi2016}, dispersive coupling between quantum systems \cite{Li2022,LachanceQuirion2020,Wolski2020} or non-reciprocal transmission of rf signals \cite{Wang2021,Zhu2020,Zhang2020,Wang2019}.

Electromagnetic cavities and ferromagnets are different versions of an harmonic oscillator, whose bosonic quanta of excitations correspond to photons and magnons, respectively. The resonant frequency of such cavities can be tuned by different means. This is relatively easy in the case of photons. For instance, size determines the fundamental modes in three dimensional microwave cavities and superconducting coplanar waveguide (CPW) resonators \cite{Pozar2011}. Superconducting thin-films are also used for the implementation of lumped element $LC$-resonators with fundamental frequencies given by $\omega_{\rm p} = 1/\sqrt{LC}$, which is, in essence, a geometrical factor as well \cite{Pozar2011}. In the case of soft magnets, spins precess around the local effective magnetic field. This field results from the contribution of the externally applied magnetic field $B_{\rm ext}$ and the (much more difficult to calculate) demagnetizing field \cite{Kittel1967}. Again, the latter is determined by the geometry of the sample. In isotropic magnets, e.g. spheres, shape effects cancel out and the resulting resonance frequency depends linearly on the external magnetic field as $\omega_{\rm m} = \gamma_e B_{\rm ext}$ \cite{Walker1957} with $\gamma_e/2\pi = 28$ GHz/T the electron gyromagnetic ratio.

In most experiments performed so far, yttrium iron garnet (YIG) spheres strongly coupled to either three dimensional cavities \cite{Tabuchi2014,Zhang2014} or superconducting CPW resonators \cite{Li2022} have been the preferred choice. These magnets are commercial and exhibit record low Gilbert damping constant ($\alpha \sim 10^{-5}$), which is beneficial for the observation of long lived coherent exchange of information between the photonic and magnonic excitations. However, YIG spheres do not couple optimally to thin film superconducting circuits that are pivotal in quantum applications \cite{Huebl2013,Blais2007}. YIG thin films are  not easily compatible with conventional lithography processes \cite{Baity2021} nor for cryogenic operation as the optimum substrate material for YIG growth becomes lossy with decreasing temperature \cite{Mihalceanu2018}. For these reasons, other C-MOS compatible materials like Permalloy (Py, with a moderate $\alpha \sim 10^{-2}$) \cite{Golovchanskiy2021,Hou2019,Li2019,Golovchanskiy2018} or iron-cobalt alloys (Fe$_{75}$Co$_{25}$, with a very promising $\alpha \sim 10^{-3}$) are drawing attention \cite{Haygood2021,Schoen2016}. 

Yet, custom tuning of the resonance frequency of such magnetic cavities is still to be demonstrated. This is relevant since experimental conditions usually impose certain constraints due to the use of, e.g., circulators or amplifiers with narrow frequency bandwidth. Additionally, operating devices at high frequencies is interesting since the coupling strength $g$ between photons and magnons increases linearly with the frequency. The reason is that $g$ is given by the Zeeman coupling between the magnon dipole moment and the oscillating vacuum field in the cavity. The latter is:
\begin{eqnarray}
b_{\rm rms}=\sqrt{ \frac{\mu_0 \hbar \omega_{\rm p}}{2 V_{\rm eff}}},
\label{F}
\end{eqnarray}
where, the sub-index denotes the root mean square (r.m.s.) and $V_{\rm eff}$ is the effective volume of the electromagnetic mode. $V_{\rm eff}$ depends on the size of the electromagnetic wavelength (thus on the inverse frequency) yielding $g \propto \omega_{\rm p}$.

Increasing $g$ is usually accomplished by a geometrical optimization of the resonator i.e., patterning of nanoconstrictions \cite{Gimeno2020,Haikka2017,Jenkins2014} and/or reducing the impedance of lumped $LC$-resonators which is equivalent to reducing $V_{\rm eff}$ \cite{Flower2019,Probst2017,Eichler2017}. These approaches serve to confine $b_{\rm rms}$ within the (typically) small volume of the ferromagnet, yielding a square-root increase of the coupling. A larger  (linear) enhancement of $g$ should be achieved by working with high frequency superconducting resonators. However, the experimental observation of such high frequency effect in isotropic ferromagnets requires polarizing the magnetization with strong  magnetic fields that are detrimental for the operation of superconducting circuits. 


Here we show a direct way to tune the resonance frequency of suitably shaped micron-sized Py samples while keeping the external magnetic field below few tens of mT.  
Using superconducting CPW transmission lines we first show that, depending on the aspect ratio of the ferromagnet, the broadband ferromagnetic resonance (FMR) spectra can be modified at will. We then convert the transmission lines into cavities by opening gap capacitors. We demonstrate that both approaches can be used to experimentally estimate the coupling factors.
Our experiments are complemented with a general-purpose theoretical model including realistic electromagnetic and micromagnetic simulations  that account for the non-homogeneous spin precession along the volume of the ferromagnet.

\begin{figure}[!ht]
\includegraphics[width=1\columnwidth]{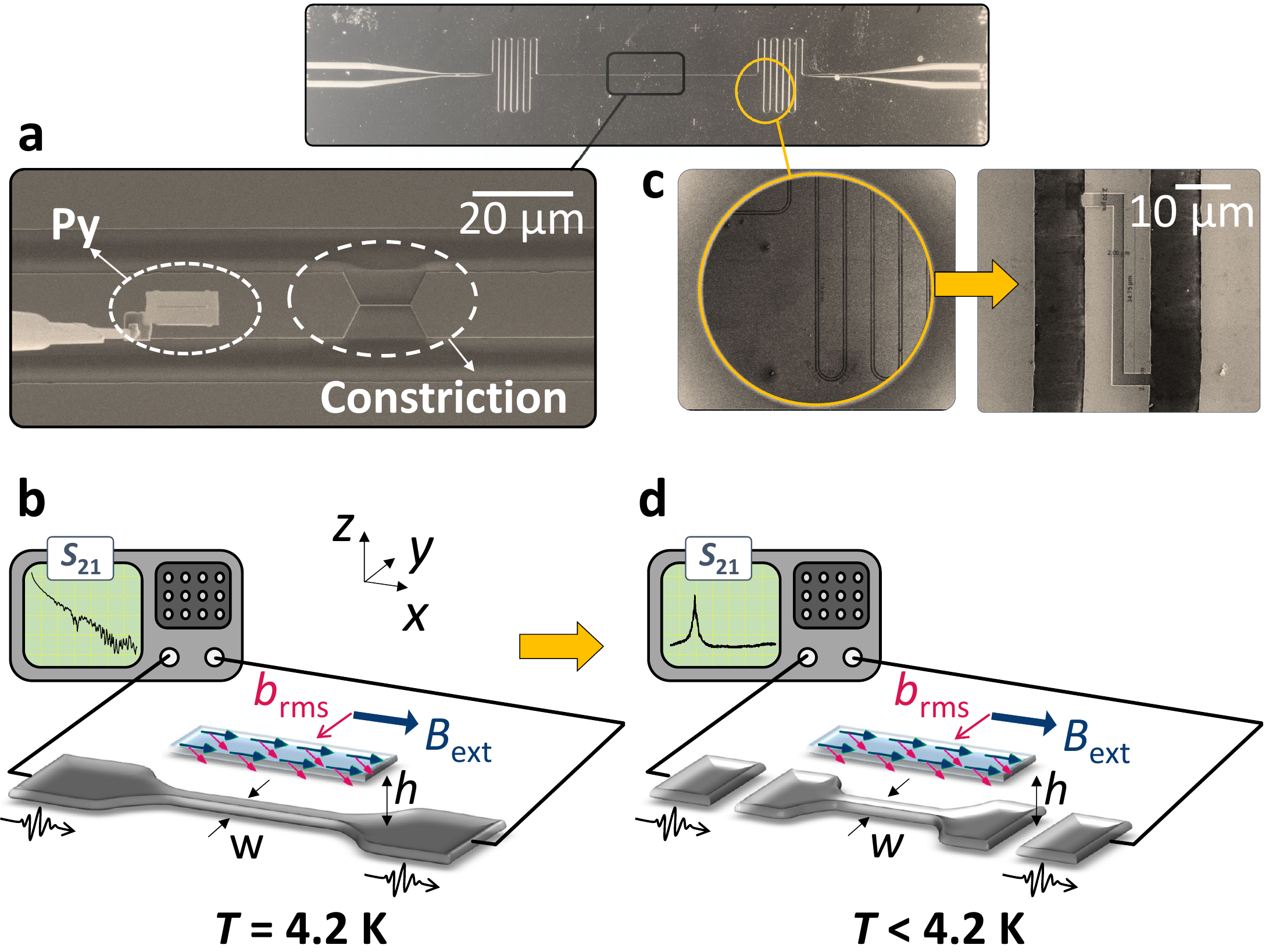}
\caption{\textbf{a}: A constriction of width $w$ is FIB patterned on the central part of a conventional Nb CPW transmission line. The Omniprobe needle is used to transport the Si$_3$N$_4$ palette with the Py ferromagnet towards the constriction. \textbf{b}: Broadband FMR experiments are performed by measuring the $S_{21}$ transmission parameter with a VNA while sweeping the external magnetic field $B_{\rm{ext}}$ applied along the $\hat{x}$ direction so that the microwave field $b_{\rm{rms}}$ produced by the transmission line is perpendicular to it. In this way, a quasi homogeneous Kittel magnon mode is excited in the ferromagnet (blue rectangle with colored arrows) lying at distance $h$ from the superconductor. \textbf{c}: Enlarged view of the meander region (left panel) where finger capacitors are opened by FIB (right panel). \textbf{d}: The resulting CPW resonator, with characteristic frequency $\omega_{\rm p}$, and its coupling to the ferromagnetic sample are characterized using the VNA.}
\label{Fig1}
\end{figure}

\section{Results}

\subsection{Sample fabrication}

We use four CPW superconducting transmission lines fabricated by optical lithography on $150$ nm thick Niobium films deposited by sputtering onto single crystalline sapphire substrates (see Fig. \ref{Fig1}\textbf{a} top). 
A constriction of width $w$ is opened in the central part of two transmission lines by focused ion beam (FIB) milling using moderate currents which result in lateral resolutions below 10 nm (see Fig. \ref{Fig1}\textbf{a} bottom). 
These constrictions, with dimensions matching those of the different samples, serve to locally focus $b_{\rm rms}$, yielding enhanced coupling with the smallest ferromagnets. 

\begin{figure*}[t]
\includegraphics[width=0.7\textwidth]{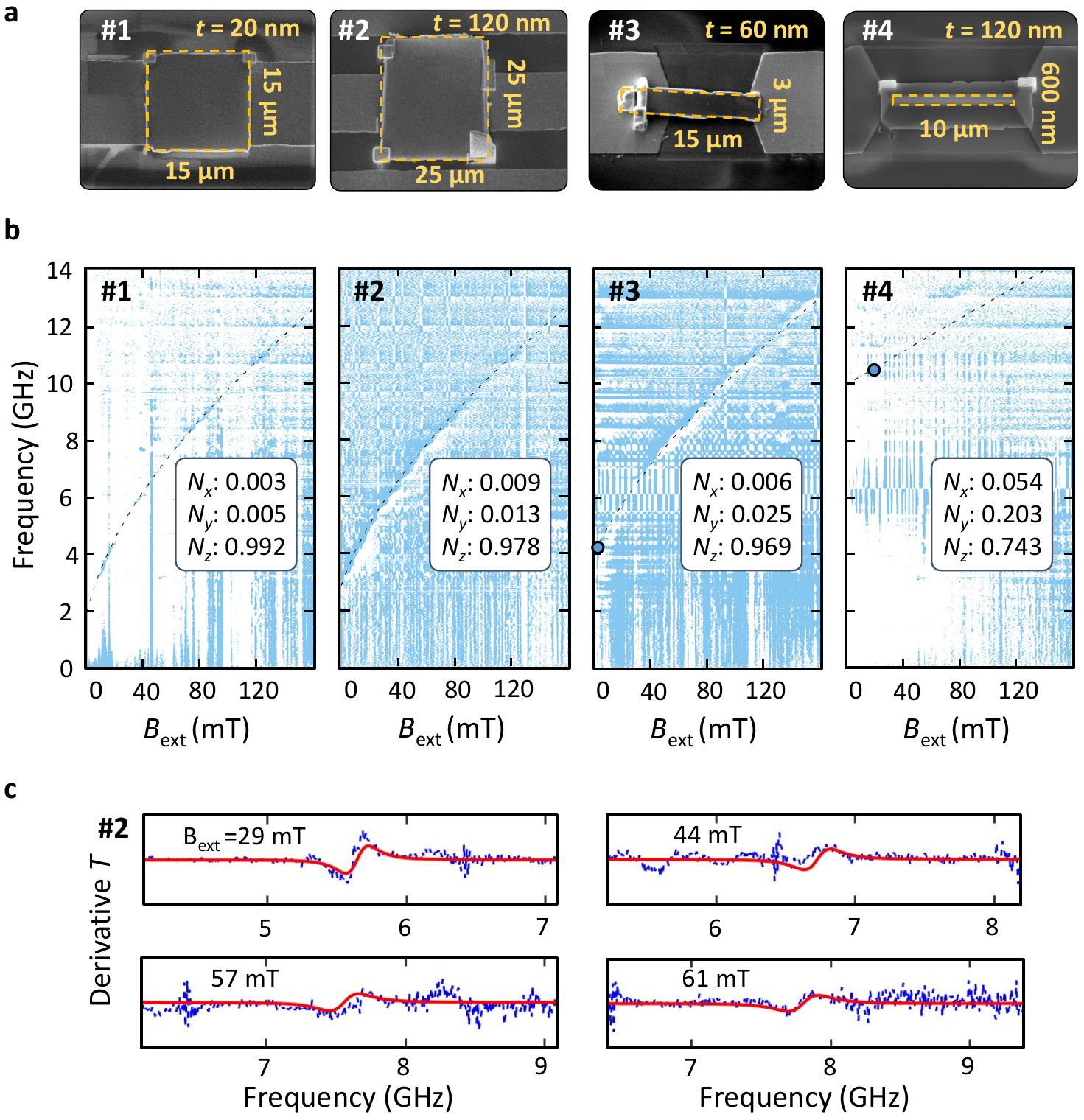}
\caption{\textbf{a}: SEM pictures of samples \textbf{\#1}, \textbf{\#2}, \textbf{\#3} and \textbf{\#4}, highlighted with yellow dashed polygons, placed onto the central line of CPW superconducting transmission lines. 
\textbf{b}: Broadband FMR curves  of the same samples in derivative mode (see \textbf{Methods}). Fitting the magnon resonance frequency $\omega_{\rm m}$ to the general Kittel formula (Eq. \ref{kittel}, dashed) yields the corresponding demagnetizing factors (legend). Color scale in all panels goes from $-0.1$ to $0.1$.
The solid dots in panels \textbf{\#3} and \textbf{\#4} correspond to the experimental data points obtained from cavity experiments (see Fig. \ref{Fig3}\textbf{b}).
\textbf{c}: Field derivative of the experimental transmission parameter (see \textbf{Methods}) corresponding to sample \textbf{\#2} (dashed blue) measured at different values of $B_{\rm ext}$ xand fits based on Eq. \ref{T} (solid red). Vertical scale goes from $-1\times10^{-3}$ to $1\times 10^{-3}$ in all panels.}
\label{Fig2}
\end{figure*}

On the other hand, Py thin-films of various thicknesses $t$ are e-beam evaporated onto $500$ nm-thick Si$_3$N$_4$ membranes. The ferromagnets are patterned by FIB milling to produce four different samples labelled  \textbf{\#1} to \textbf{\#4} whose dimensions are summarized in Fig. \ref{Fig2}\textbf{a}. Finally, an Omniprobe needle inside the FIB microscope is used to transport a micrometric sized Si$_3$N$_4$ palette containing the Py sample to precise positions on top of the superconducting circuit reaching optimum coupling \cite{MartinezPerez2018a}. For this purpose, it is important to minimize the distance $h$ between the ferromagnet and the superconducting line (Fig. \ref{Fig1}\textbf{b}). This is difficult in case of sample \textbf{\#2} due to its large size that causes the Si$_3$N$_4$ palette to bend. We estimate $h \sim 4$ $\mu$m for sample \textbf{\#2} and $h\sim500$ nm for the rest. Samples \textbf{\#1} and \textbf{\#2} are located onto the as-fabricated transmission lines, sample \textbf{\#3} goes onto the $w=1$ $\mu$m wide constriction and, finally, sample \textbf{\#4} lies on top of the smallest $w=240$ nm wide constriction (see Fig. \ref{Fig2}\textbf{a}.

\subsection{Broadband FMR}

Samples are cooled down to 4.2 K in liquid helium in a cryostat containing a superconducting vector magnet. Broadband FMR experiments are performed by measuring the transmission coefficient $S_{\rm 21}$ through the transmission line using a vector network analyzer (VNA) as schematized in  Fig. \ref{Fig1}\textbf{b}. The external magnetic field $B_{\rm ext}$ is used to polarize the ferromagnet parallel to the transmission line ($\hat x$ direction) so that the microwave field produced by the superconductor (along the $\hat y$ direction) can excite the homogeneous Kittel mode with frequency given by \cite{Kittel1967}:
\begin{eqnarray}
\begin{split}
 \omega_{\rm m}^2= \gamma_e^2 \Big[B_{\rm ext} + & (N_y-N_x)\mu_0 M_{\rm s}\Big] \times \\ & \Big[B_{\rm ext}+(N_z-N_x)\mu_0 M_{\rm s}\Big].
 \label{kittel}
 \end{split}
\end{eqnarray}
Here, $\mu_0 = 4\pi \times 10^{-7}$ Tm/A and $M_{\rm s} = 0.86 \times 10^6$ A/m is the saturation magnetization of Py measured in a nominally identical substrate. Finally, $N_x$, $N_y$ and $N_z$ are the magnetometric demagnetizing factors that satisfy $N_x+N_y+N_z=1$. These factors can be analytically estimated for each sample based on geometrical considerations \cite{Aharoni1998}. Results are given in Table \ref{tableN}.

\begin{table}[t]
	\begingroup

	\def\arraystretch{1.2}
	\begin{tabular}{ p{0.7cm}| p{1cm} p{1cm} p{1cm} | p{1cm} p{1cm} p{1cm} }
		\hline   
       & \multicolumn{3}{c|}{Calculated}  & \multicolumn{3}{c}{ Fitted } \\
			 & $N_x$ & $N_y$& $N_z$             & $N_x$& $N_y$& $N_z$          \\
		\hline
		\textbf{\#1}  & 0.003  & 0.003 & 0.994 & 0.003 & 0.005 & 0.992\\
		\textbf{\#2}  & 0.009  & 0.009 & 0.982 & 0.009 & 0.013 & 0.978 \\
		\textbf{\#3}  & 0.006  & 0.033 & 0.960 & 0.006 & 0.025 & 0.969 \\
		\textbf{\#4}  & 0.011  & 0.195 & 0.795 & 0.054 & 0.203 & 0.743\\
		\hline  
	\end{tabular}
	\caption{
Demagnetizing factors calculated  from geometrical considerations \cite{Aharoni1998} and fits based on Eq. \ref{kittel}.
	}
	\label{tableN}
	\endgroup
\end{table}

Experimental curves corresponding to samples \textbf{\#1}, \textbf{\#2}, \textbf{\#3} and \textbf{\#4} are shown in Fig. \ref{Fig2}\textbf{b}. The Kittel resonance frequency depends strongly on the aspect ratio of the Py sample. As a matter of fact, $\omega_{\rm m}$ increases from 2 GHz up to more than 10 GHz at $B_{\rm ext} \sim 0$ mT.  Eq. \ref{kittel} accounts well for all experimental curves, allowing to fit the demagnetizing factors 
(see insets of Fig. \ref{Fig2}\textbf{b} and Table \ref{tableN}).   In case of samples  \textbf{\#1} and \textbf{\#2}, symmetry considerations would yield $N_x=N_y$. However, this is not the case due to unavoidable imperfections that are especially dramatic in case of sample \textbf{\#2} stemming from the above mentioned  bending. In case of sample  \textbf{\#3} and, especially,  \textbf{\#4}, $N_y$ increases considerably, indicating that the $\hat y$ direction becomes progressively harder.
 As expected, $N_x$ increases slightly with sample thickness. In case of sample \textbf{\#4} we found $N_x$ almost five times larger than the calculated value. This is probably due to a small missalignment of the external magnetic field with respect to the sample's easy axis. This effect is much more critical in samples with very large aspect ratio as \textbf{\#4}.

FMR data allow us to estimate the frequency-dependent magnon-photon coupling. From input-output theory, we can express the transmission through an open line as \cite{Roy2017}:
\begin{eqnarray}
\label{tw}
    T= 
    1 - 
     \frac{\Gamma}{\Gamma +  \gamma + {\mathrm i} (\omega_{\rm m} - \omega)}.
\label{T}
\end{eqnarray}
We recall that $\omega_{\rm m}$ depends on the magnetic field through Eq. \ref{kittel}. In the above formula, $\gamma = 2 \alpha \omega_{\rm m}$ is the linewidth of the ferromagnetic resonance and $\Gamma = g^2 \pi/\omega_{\rm m}$.

We use Eq. \ref{T} to fit the experimental data (in derivative mode, as explained in the \textbf{Methods} section). A few representative curves corresponding to sample \textbf{\#2} are shown in Fig. \ref{Fig2}\textbf{c}. From these fits we estimate $\alpha \sim 0.01$, i.e.,  the Py damping factor at cryogenic temperatures. This value is close to the Gilbert damping parameter measured at low and room temperatures  \cite{Zhao2016}. On the other hand, we can estimate the frequency-dependent coupling $g_{\rm FMR}$ where the sub-index is used to indicate that these values are obtained from FMR experiments. Table \ref{table} gives coupling values determined at some fixed frequencies $\omega_{\rm m}=\omega_{\rm p}$.

\subsection{Cavity experiments}

We next transform each transmission line into a resonator by opening gap capacitors using FIB milling (see Fig. \ref{Fig1}\textbf{c}). In doing so, we chose the resulting cavity length to fit the $\lambda /2$ mode of the desired frequency $\omega_{\rm p}$ summarized in Table \ref{table}. Also, the capacitors are designed to yield resonators with moderate quality factor $Q \sim 1 - 2 \times 10^3$ which are easier to measure while keeping $\kappa < \gamma$ \cite{Goeppl2008}. Samples are again immersed in liquid helium except for sample \textbf{\#4} that is cooled down to $10$ mK using a dilution refrigerator. A VNA is used to probe the $S_{\rm 21}$ factor as a function of frequency while sweeping $B_{\rm ext}$ as described in the transmission line experiments (see Fig. \ref{Fig1}\textbf{d}). 

\begin{figure}[t]
	\includegraphics[width=1\columnwidth]{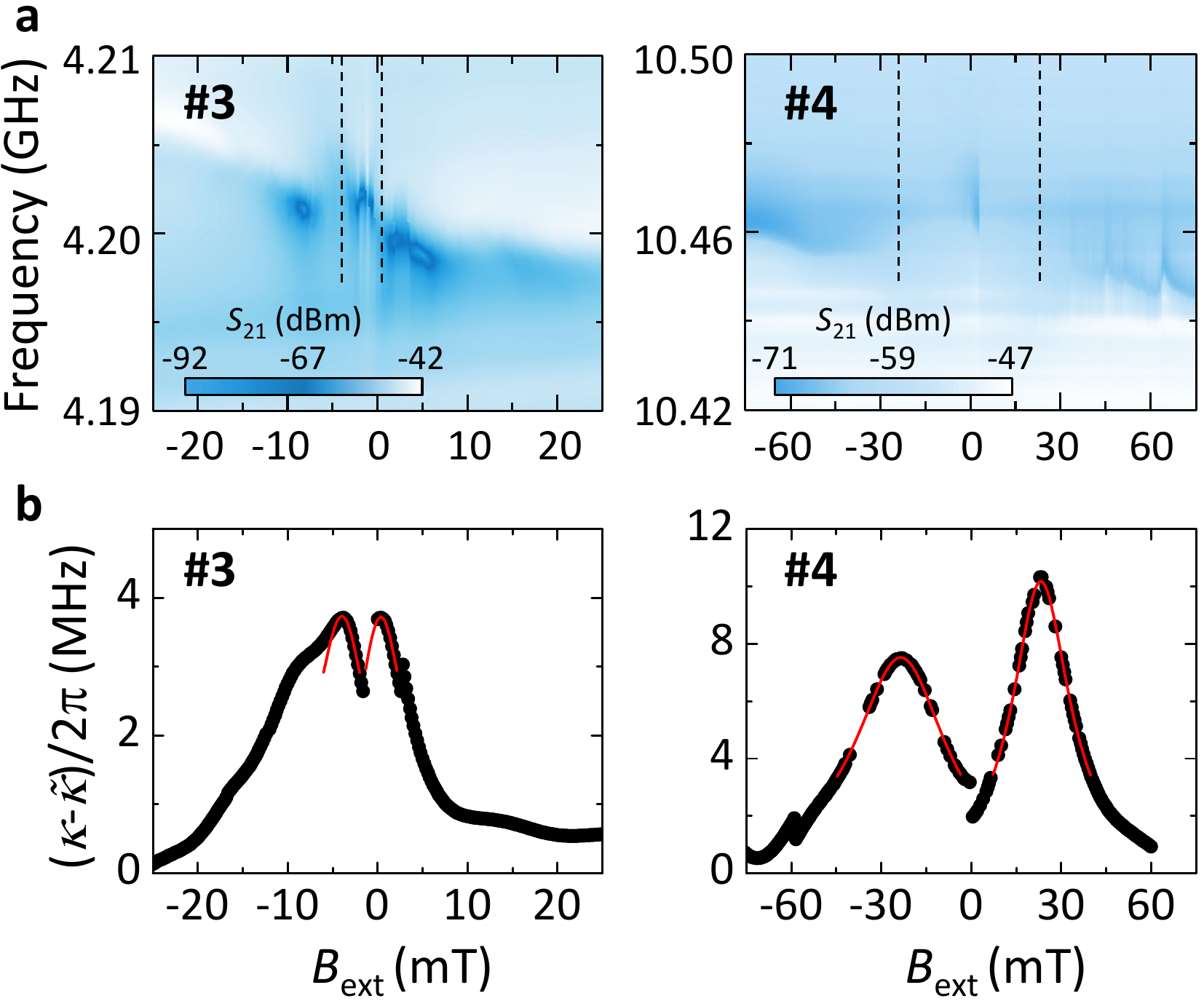}
	\caption{\textbf{a}: Experimental microwave transmission through superconducting resonators containing samples \textbf{\#3} ($\omega_{\rm p}/2\pi=4.2$ GHz, $T=4.2$ K) and \textbf{\#4} ($\omega_{\rm p}/2\pi=10.5$ GHz, $T=10$ mK). The dashed lines are the broadband FMR curves shown in Fig. \ref{Fig2}\textbf{b}. \textbf{b}: Field dependence of the resonance width $\kappa$ for the same resonators and least-squares fits (dashed lines) to Eq. \ref{kappa}. Data corresponding to sample \textbf{\#3} are shifted 2 mT towards negative fields due to a small remnant field in the superconducting coils. Maxima of $\kappa$ vs. $B_{\rm ext}$ provide two additional experimental points of FMR at $B_{\rm ext} =-3.9$ and $0.4$ mT for sample \textbf{\#3} and  $B_{\rm ext} =\pm 23$ mT for sample  \textbf{\#4} (see Fig. \ref{Fig2}\textbf{b}).}
	\label{Fig3}
\end{figure}

Measurements performed on \textbf{\#1} and \textbf{\#2} evidence that these samples are in the weak coupling regime. On the other hand, two avoided crossings can be distinguished in the experimental data obtained with samples \textbf{\#3} and \textbf{\#4} (see Fig. \ref{Fig3}\textbf{a}). These anticrossings appear at positive and negative magnetic fields, at the intersect with the FMR curves (dashed lines), which is a proof of magnon-photon coupling. In case of sample \textbf{\#4}, we observe also a number of small successive anticrossings at positive fields. The origin of such features is unknown but they might arise from the interaction between the fundamental mode of the resonator and spurious modes related with the magnetic behavior of the superconducting cavity itself, i.e., they are not related with the Permalloy sample.

\begin{table}[t]
	\begingroup
	\setlength{\tabcolsep}{3pt} 
	\def\arraystretch{1.2}
	\begin{tabular}{r|ccccc}
		\hline   
		   & $\omega_{\rm p}/2\pi$  & $g_{\rm FMR}/2\pi$  & $g_{\rm res}/2\pi$  & $g_{\rm theo}/2\pi$ \\
     & (GHz)  & (MHz) & (MHz)  & (MHz)    \\
		\hline
		\textbf{\#1}   & 2.8  & 7 & - & 6.6  \\
		\textbf{\#2} & 3.5   &16 & - & 17.3  \\
		\textbf{\#3}  & 4.2  &36 & 36 & 45.0  \\
		\textbf{\#4} & 10.5   & 66 & 72 & 72.4 \\
		\hline  
	\end{tabular}
	\caption{
	Photon resonance frequency ($\omega_{\rm p}$); photon-magnon coupling determined  from FMR measurements ($g_{\rm FMR}$), from cavity experiments ($g_{\rm res}$) and from theory ($g_{\rm theo}$). The latter is calculated for $h=4$ $\mu$m in case of sample \textbf{\#2} and $h=500$ nm for the rest. 
	}
	\label{table}
	\endgroup
\end{table}


Cavity experiments can be also used to give an alternative estimate of the magnon-photon coupling. For this purpose we analyze the cavity resonance linewidth $\kappa$ as a function of external magnetic field. The linewidth is evaluated by fitting the experimental transmission to a Fano-like resonance (see \textbf{Methods}). The latter accounts better for asymmetries in our experimental data that likely arise from interference between the resonator and a small continuous background signal \cite{Fano1961}. Depending on the coupling strength, the resonance broadens around the intersection field according to \cite{Bushev2011}:
\begin{eqnarray}
\label{kappa}
    \kappa= 
    \tilde{\kappa} + 
     \frac{\gamma g^2}{(\omega_{\rm p} - \omega_{\rm m})^2+\gamma^2},
\end{eqnarray}
where $\tilde{\kappa}$ is the linewidth of the bare cavity that can be measured far from resonance. Again, we recall that $\omega_{\rm m}$ depends on the external field through Eq. \ref{kittel}. Experimental $\kappa$ vs. $B_{\rm ext}$ curves exhibit two clear maxima at positive and negative fields, where level repulsion occurs (Fig. \ref{Fig3}\textbf{b}). Fitting these data to Eq. \ref{kappa}  gives a complementary estimate of the coupling (labelled $g_{\rm res}$) that we summarize in Table \ref{table}.

\subsection{Theoretical model}

We finish by comparing our experimental results with numerical estimations of the coupling factors. 
A generalization of formulas 
given in Refs. \onlinecite{MartinezPerez2019,MartinezPerez2018},  where the field was considered homogeneous in the relevant region, follows.
Each ferromagnetic sample is discretized into cells.
The Hamiltonian interaction is given by $H_{\rm I}= v_{\rm cell} \sum_{j} {\bf m}_j\cdot {\bf b}_{\rm rms} ( {\bf r}_j)$.
Here, $v_{\rm cell}$ and ${\bf m}_j$ are the volume and   magnetization  of the cell, respectively, while ${\bf b}_{\rm rms} ({\bf r } _j )$ is the rms vacuum field at the cell's position ${\bf r } _j$.
In each cell, the number of spins is large enough to perform the Holstein-Primakoff transformation, so the quantization of the magnetization can be written in terms of bosonic operators as ${\bf m}_j = \Delta {\bf M}_j (a_j ^\dagger + a_j)$ where
$\Delta {\bf M}_j$ is the position-dependent amplitude of the magnetization modulation. The latter will be always valid as we are working in the low photon or low power limit. 
Once this is done, we define  the \textit{collective} mode \cite{Huemmer2012}:
$$
    a= \frac{1}{\sqrt{ \sum_j  \left (\Delta {\bf M}_j \cdot {\bf b}_{\rm rms} ({\bf r}_j) \right )^2} }  \sum_j \Delta {\bf M}_j \cdot {\bf b}_{\rm rms} ({\bf r}_j) \;  {a_j}
$$

Finally, taking the continuum limit, the magnon-resonator coupling reads: 
\begin{eqnarray}
g^2=\frac{ \gamma v_{\rm cell} }{4 \hbar} \sum_j \textbf{b}_{\rm rms}({\bf r}_j) \cdot \Delta \textbf{M}_j  ,
\label{gteo}
\end{eqnarray}
which is valid for \textit{any} kind of magnonic mode excited in the ferromagnet, including the Kittel mode studied here. 

\begin{figure}[t]
	\includegraphics[width=1\columnwidth]{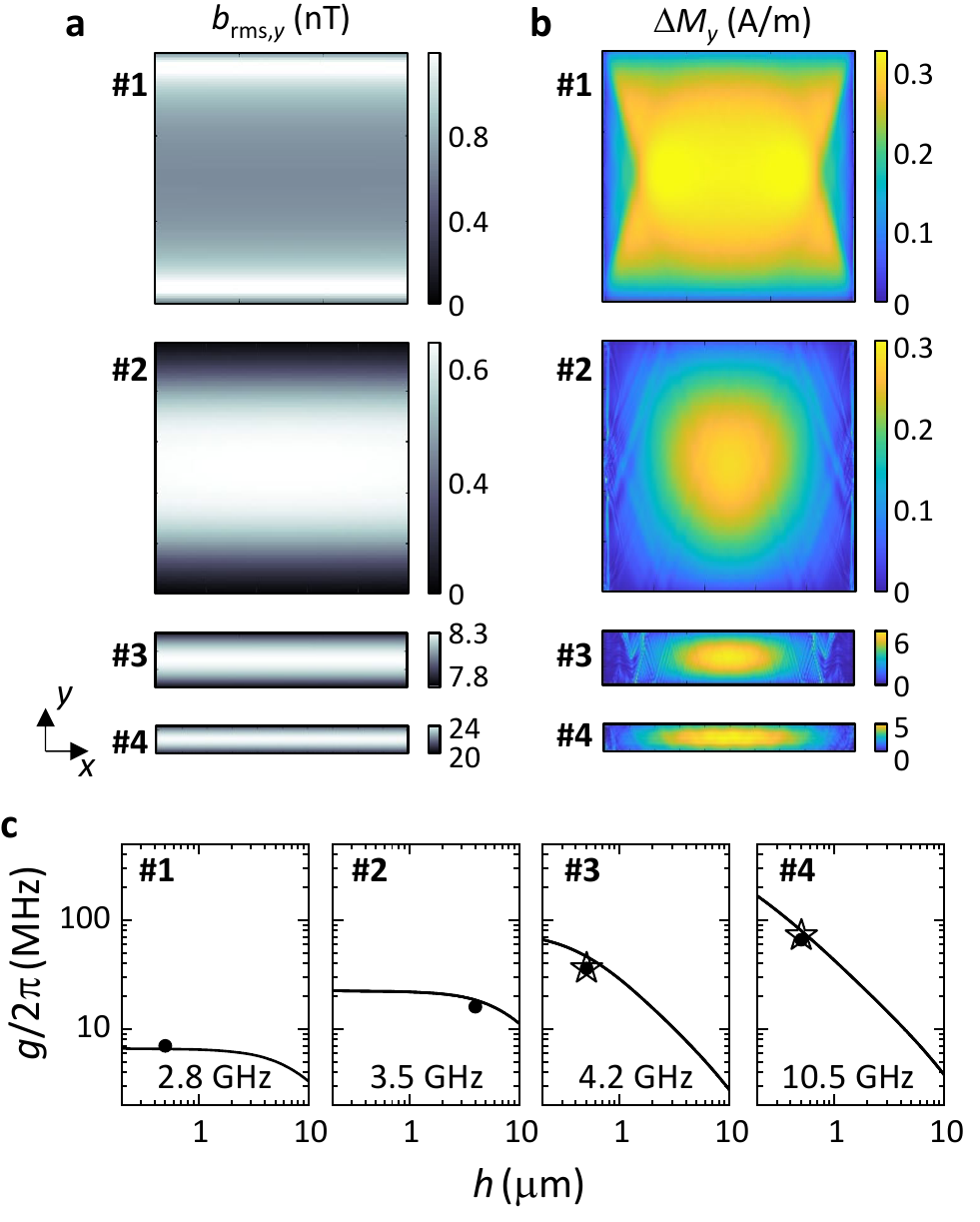}
	\caption{ Spatial distribution of the in-plane ($y$) component of $b_{\rm rms}$ (\textbf{a}) and $\Delta {M}$ (\textbf{b}) calculated at the position of the ferromagnet. \textbf{c}: Solid lines are the numerical  $g_{\rm theo}$ (at fixed frequency $\omega_{\rm p}$ given above each panel) vs. distance $h$.  
	Data points are the experimental $g_{\rm FMR}$ (dots) and $g_{\rm res}$ (stars). From left to right, our experimental data roughly follow $g \propto \omega_{\rm p}$.
}
	\label{Fig4}
\end{figure}

We numerically calculate the spatial distribution of $\textbf{b}_{\rm rms}({\bf r}_j)$ for the different resonators used in this work (see \textbf{Methods}). Figure \ref{Fig4}\textbf{a} shows the resulting in-plane ($y$) component of $b_{\rm rms}$ for the different samples obtained at the position of the ferromagnet. We highlight that $b_{\rm rms}$ increases by a factor $\sim 30$ from sample \textbf{\#1} to \textbf{\#4} stemming from the smallness of the patterned constriction.  

On the other hand, we need to estimate  $\Delta \textbf{M}_j  $ which depends on the spatial distribution of the excitation field, i.e., $\textbf{b}_{\rm rms}({\bf r}_j)$, and also on the non-homogeneous magnetic susceptibility of the ferromagnet. Figure \ref{Fig4}\textbf{b} shows the numerically calculated $y$ component of $\Delta {M}$ for the different samples (see \textbf{Methods}). In general, $\Delta M_y$ is maximum in the inner part of the ferromagnets but decreases considerably when approaching the edges. This effect is less pronounced in case of sample \textbf{\#1}, for which the susceptibility is substantially more homogeneous. This is due to its small thickness that constrains the modulation of the magnetization to the plane.

Finally, we use Eq. \ref{gteo} to calculate the theoretical coupling factors $g_{\rm theo}$ at fixed frequency $\omega_{\rm p}$ for varying distance between the ferromagnet and the superconducting line, i.e., distance $h$ in Fig. \ref{Fig1}\textbf{b} and \textbf{d}. Results are plotted in Fig. \ref{Fig4}\textbf{c} (solid lines) together with the experimental $g$ values obtained at the same frequencies (scatter). 

\section{Discussion}
  
We start by highlighting that, in case of samples \textbf{\#3} and \textbf{\#4}, $g_{\rm FMR} \approx g_{\rm res}$. This demonstrates that the two experimental approaches, i.e., transmission and cavity configuration, are equivalent and valid to estimate the coupling. This result is not limited to the use of ferromagnetic samples but applies to whatsoever hybrid light-matter architecture based on, e.g., paramagnetic spins or superconducting qubits.

We can calculate the effective coupling per spin for samples \textbf{\#3} and \textbf{\#4} that amounts to $70$ and $280$ Hz, respectively. The latter 
approaches the largest spin-photon coupling strength achieved so far in optimized superconductor-peramalloy-insulator hybrid samples, i.e., 350 Hz \cite{Golovchanskiy2021a}
and exceeds the largest spin-photon coupling measured with paramagnetic spins under similar conditions \cite{Gimeno2020}. As a matter of fact, the $g$ values reported here are similar to those obtained using optimized $LC$-resonators \cite{Probst2017, Hou2019}. 

We next focus our attention on the experimental $g_{\rm FMR}$ and $g_{\rm res}$ and compare them to the $g_{\rm theo}$ calculated with our theory (Table \ref{table}). The main source of error in $g_{\rm theo}$ is the distance $h$. As shown in Fig. \ref{Fig4}\textbf{c}, experiment and theory agree to a very good extent for the estimated values of $h$, confirming the validity of our model. 

Estimating the coupling strength in hybrid magnon-photon architectures is essential for designing reliable experiments. Here, we exploit the shape of thin-film ferromagnets as tuning knob to modify their resonance frequency. This requires an additional account for the largely non-uniform demagnetizing field in rectangular prisms.  However, little attention has been given till now to the distribution of magnetization precession along the volume of real ferromagnets \cite{Macedo2021, Bourhill2020}.  In contrast, our approach involves the calculation of both the three-dimensional distribution of vacuum field created by the cavity and the magnetic susceptibility of the ferromagnet. This is important even for Kittel modes in ellipsoids of revolution but becomes critical when dealing with realistic thin-films for which magnetization precession might be highly elliptical. Examples of the latter include spin modes in rectangular prisms but also skyrmions or flux-closure magnetic textures like vortices \cite{MartinezPerez2018,MartinezPerez2019}. As we have demonstrated,
 Eq. \ref{gteo} can be used for calculating the coupling between superconducting circuits of whatever geometry and magnetostatic modes of any kind excited in ferromagnetic samples of arbitrary shape as long as this mode can be efficiently reproduced by micromagnetic simulations.


Finally, our experimental data serve to demonstrate the direct relation between the coupling strength and the cavity frequency (see Fig. \ref{Fig4}\textbf{c}). Deviations from the expected linear dependence arise from the different geometries of the resonators  and the particular volumes and demagnetizing factors of each sample. The latter affects the magnetic susceptibility of the corresponding magnetostatic (Kittel) mode. Besides these factors, our experimental data roughly follow $g \propto \omega_{\rm p}$. This is particularly evident for samples \textbf{\#3} and \textbf{\#4}. Increasing the cavity operation frequency from $\omega_{\rm p}/2\pi = 4.2$ GHz (\textbf{\#3}) up to $\omega_{\rm p}/2\pi = 10.5$ GHz (\textbf{\#4}) yields twice the coupling strength.   

We stress that patterning nanoconstrictions or reducing the impedance yields a square-root increase of $g$ while keeping the losses in the ferromagnet unchanged, meaning these approaches serve to increase the cooperativity. On the other hand, increasing the operation frequency yields a linear increase of $g$, at the cost of also increasing $\gamma = 2 \alpha \omega_{\rm p}$. That is to say, increasing $\omega_{\rm p}$ does not ease entering into the strong coupling regime but does have the effect of linearly increasing the rate of information exchange between photon and magnon excitations.


\section{Conclusions}

We demonstrate the possibility of tuning the operation frequency in cavity magnonics by controlling the shape of the ferromagnet. Varying the aspect ratio of different Py samples we reach operation frequencies up to $10.5$ GHz at $\sim 20$ mT. Working at such high frequencies with isotropic spherical ferromagnets would have required an external field close to $0.4$ T. High frequency operation together with the use of patterned nanoconstrictions in superconducting transmission lines enhances the net coupling between magnetic materials and cavity photons allowing us to reach an average single spin coupling of $280$ Hz.

Our experimental results demonstrate that the light-matter coupling strength can be reliably estimated from both open transmission or cavity experiments. Besides, this has allowed combining broadband spectroscopic measurements, which provide the full picture of the ferromagnetic excitations, with experiments on superconducting resonators, which show how these modes couple to cavity photons. 

Finally, we provide a simple recipe to theoretically calculate $g$ in hybrid systems combining superconducting circuits and  \textit{realistic} ferromagnetic materials.  Our approach can be used to design quantum magnonic experiments exploiting not only the homogeneous Kittel mode but any higher order magnonic mode in ferromagnets of arbitrary size and shape and to tune the operation conditions, frequency and magnetic field, to fit those required in a particular experiment or application.


\section{Methods}

\subsection{Data fitting}

Open transmission data is analyzed by performing the pseudo-derivative of the transmission parameter. For this purpose, we  calculate 
$$
\text{Derivative} \; T_{\rm exp} = \frac{T_{\rm exp}(B_{\rm ext}^{i+1})-T_{\rm exp}(B_{\rm ext}^i)}{T_{\rm exp}(B_{\rm ext}^{i+1})},
$$
where $T_{\rm exp}$ is the linear amplitude of the transmission $T_{\rm exp} = 10^{S_{\rm 21}/20}$ and super-indexes $i$ and $i+1$ refer to successive field values. Resulting curves vs. frequency and field are ploted in Fig. \ref{Fig2}\textbf{b} and \textbf{c}. We then calculate the pseudo-derivative of the absolute value of Eq. \ref{T}, which is given by:
$$
|T|=\sqrt{ \frac{\gamma^2 +(\omega_{\rm m} - \omega)^2}{\Gamma^2 + 2 \Gamma \gamma + \gamma^2 + (\omega_{\rm m} - \omega)^2} },
$$
where $\omega_{\rm m}$ and $\gamma = 2 \alpha \omega_{\rm m}$ are evaluated at each $B_{\rm ext}^i$ using Eq. \ref{kittel}. We finally fit the pseudo-derivative of $|T|$ to the experimental data as shown in Fig. \ref{Fig2}\textbf{c} to obtain the coupling strenght $g=(\Gamma \omega_{\rm m} / \pi )^{1/2}$.

Cavity experiments are analized by fitting the experimental resonance to a Fano curve:
$$
T = y_0 + A \frac {\big(  \frac{\omega-\omega_0}{\gamma/2} + q \big)^2}{ 1 + \big(\frac{\omega-\omega_0}{\gamma/2}\big)^2} \frac{1}{1+q^2},
$$
with $y_0$ an offset, $A$ the amplitude and $q$ the Fano parameter that defines the lineshape of the resonance. Finally, $1/(1+q^2 )$ is a normalization factor. Fano resonances can be observed when the resonators is not perfectly matched to the input output lines, yielding a parasitic path between connectors. $q = 0$ yields an antiresonance whereas for $q \rightarrow \pm \infty$ the Lorentzian resonance with linewidth $\gamma$ is recovered.

\subsection{simulations}

The spatial distribution of $\textbf{b}_{\rm rms}({\bf r}_j)$ is calculated using the finite-element software 3D-MLSI \cite{Khapaev2002} that  solves London equations in a user-defined superconducting circuit with flowing current $i_{\rm rms}$. The latter is the rms zero-point current  that can be obtained as \cite{Jenkins2013}:
\begin{eqnarray}
i_{\rm rms}=\omega_{\rm p}\sqrt{\frac{\hbar\pi}{4 Z_0}},
\label{irms}
\end{eqnarray}
with $Z_0=50$ $\Omega$ the impedance of the circuit. Inserting the experimental frequencies $\omega_{ \rm p}$ into Eq. \ref{irms}, we calculate $i_{\rm rms}$ for each resonator. From here, 3D-MLSI allows calculating the spatial distribution of supercurrents and the resulting $\textbf{b}_{\rm rms}({\bf r}_j)$ at each cell $j$ inside the volume of the ferromagnet. 

We finally calculate the spatial distribution of $\Delta \textbf{M}_j  $ resulting from the spatial-dependent $\textbf{b}_{\rm rms}({\bf r}_j)$ using the micromagnetic solver MUMAX$^3$ \cite{Vansteenkiste2014}. For this purpose, we excite the ferromagnet with a time- and spatial-dependent field $\textbf{b}_{\rm excit} = \textbf{b}_{\rm rms}({\bf r}_j) \cos(\omega_{\rm m} t)$. As a result, we obtain the time evolution of the magnetization at each cell $j$ and fit the resulting curve to an equation of the form $\Delta \textbf{M}_j \sin (\omega_{\rm m} t + \delta_j)$. The coupling is finally calculated applying Eq. \ref{gteo}.


\section{Acknowledgments}
\label{acknowledgments}

This work  was partly funded and supported by the spanish MCIN/ AEI /10.13039/501100011033/ and FEDER (\textit{Una manera de hacer Europa}) through grant RTI2018-096075-B-C21, the BBVA Foundation through "Beca Leonardo a Investigadores y Creadores Culturales 2019", the Arag\'on Regional Government through project E09$\_$20R (\textit{Construyendo Europa desde Arag\'on}), CSIC through project 202160I034, the CSIC program for the Spanish Recovery, Transformation and Resilience Plan funded by the Recovery and Resilience Facility of the European Union, established by the Regulation (EU) 2020/2094, the EU through FET-OPEN (862893 FATMOLS) and the European Research Council (ERC) under the European Union’s Horizon 2020 research and innovation programme (948986 QFaST). S. M.-L. R. acknowledges a FPI grant from the spanish MCIN. Authors would like to acknowledge the use of Servicio General de Apoyo a la Investigaci\'on-SAI, Universidad de Zaragoza.

\bibliographystyle{apsrev4-2}
\bibliography{sergio}

\begin{thebibliography}{49}%
\makeatletter
\providecommand \@ifxundefined [1]{%
 \@ifx{#1\undefined}
}%
\providecommand \@ifnum [1]{%
 \ifnum #1\expandafter \@firstoftwo
 \else \expandafter \@secondoftwo
 \fi
}%
\providecommand \@ifx [1]{%
 \ifx #1\expandafter \@firstoftwo
 \else \expandafter \@secondoftwo
 \fi
}%
\providecommand \natexlab [1]{#1}%
\providecommand \enquote  [1]{``#1''}%
\providecommand \bibnamefont  [1]{#1}%
\providecommand \bibfnamefont [1]{#1}%
\providecommand \citenamefont [1]{#1}%
\providecommand \href@noop [0]{\@secondoftwo}%
\providecommand \href [0]{\begingroup \@sanitize@url \@href}%
\providecommand \@href[1]{\@@startlink{#1}\@@href}%
\providecommand \@@href[1]{\endgroup#1\@@endlink}%
\providecommand \@sanitize@url [0]{\catcode `\\12\catcode `\$12\catcode
  `\&12\catcode `\#12\catcode `\^12\catcode `\_12\catcode `\%12\relax}%
\providecommand \@@startlink[1]{}%
\providecommand \@@endlink[0]{}%
\providecommand \url  [0]{\begingroup\@sanitize@url \@url }%
\providecommand \@url [1]{\endgroup\@href {#1}{\urlprefix }}%
\providecommand \urlprefix  [0]{URL }%
\providecommand \Eprint [0]{\href }%
\providecommand \doibase [0]{https://doi.org/}%
\providecommand \selectlanguage [0]{\@gobble}%
\providecommand \bibinfo  [0]{\@secondoftwo}%
\providecommand \bibfield  [0]{\@secondoftwo}%
\providecommand \translation [1]{[#1]}%
\providecommand \BibitemOpen [0]{}%
\providecommand \bibitemStop [0]{}%
\providecommand \bibitemNoStop [0]{.\EOS\space}%
\providecommand \EOS [0]{\spacefactor3000\relax}%
\providecommand \BibitemShut  [1]{\csname bibitem#1\endcsname}%
\let\auto@bib@innerbib\@empty
\bibitem [{\citenamefont {Blais}\ \emph {et~al.}(2007)\citenamefont {Blais},
  \citenamefont {Gambetta}, \citenamefont {Wallraff}, \citenamefont {Schuster},
  \citenamefont {Girvin}, \citenamefont {Devoret},\ and\ \citenamefont
  {Schoelkopf}}]{Blais2007}%
  \BibitemOpen
  \bibfield  {author} {\bibinfo {author} {\bibfnamefont {A.}~\bibnamefont
  {Blais}}, \bibinfo {author} {\bibfnamefont {J.}~\bibnamefont {Gambetta}},
  \bibinfo {author} {\bibfnamefont {A.}~\bibnamefont {Wallraff}}, \bibinfo
  {author} {\bibfnamefont {D.~I.}\ \bibnamefont {Schuster}}, \bibinfo {author}
  {\bibfnamefont {S.~M.}\ \bibnamefont {Girvin}}, \bibinfo {author}
  {\bibfnamefont {M.~H.}\ \bibnamefont {Devoret}},\ and\ \bibinfo {author}
  {\bibfnamefont {R.~J.}\ \bibnamefont {Schoelkopf}},\ }\href@noop {}
  {\bibfield  {journal} {\bibinfo  {journal} {Physical Review A}\ }\textbf
  {\bibinfo {volume} {75}},\ \bibinfo {pages} {032329} (\bibinfo {year}
  {2007})}\BibitemShut {NoStop}%
\bibitem [{\citenamefont {Harder}\ \emph {et~al.}(2021)\citenamefont {Harder},
  \citenamefont {Yao}, \citenamefont {Gui},\ and\ \citenamefont
  {Hu}}]{Harder2021}%
  \BibitemOpen
  \bibfield  {author} {\bibinfo {author} {\bibfnamefont {M.}~\bibnamefont
  {Harder}}, \bibinfo {author} {\bibfnamefont {B.~M.}\ \bibnamefont {Yao}},
  \bibinfo {author} {\bibfnamefont {Y.~S.}\ \bibnamefont {Gui}},\ and\ \bibinfo
  {author} {\bibfnamefont {C.-M.}\ \bibnamefont {Hu}},\ }\href@noop {}
  {\bibfield  {journal} {\bibinfo  {journal} {Journal of Applied Physics}\
  }\textbf {\bibinfo {volume} {129}},\ \bibinfo {pages} {201101} (\bibinfo
  {year} {2021})}\BibitemShut {NoStop}%
\bibitem [{\citenamefont {Li}\ \emph {et~al.}(2020)\citenamefont {Li},
  \citenamefont {Zhang}, \citenamefont {Tyberkevych}, \citenamefont {Kwok},
  \citenamefont {Hoffmann},\ and\ \citenamefont {Novosad}}]{Li2020}%
  \BibitemOpen
  \bibfield  {author} {\bibinfo {author} {\bibfnamefont {Y.}~\bibnamefont
  {Li}}, \bibinfo {author} {\bibfnamefont {W.}~\bibnamefont {Zhang}}, \bibinfo
  {author} {\bibfnamefont {V.}~\bibnamefont {Tyberkevych}}, \bibinfo {author}
  {\bibfnamefont {W.-K.}\ \bibnamefont {Kwok}}, \bibinfo {author}
  {\bibfnamefont {A.}~\bibnamefont {Hoffmann}},\ and\ \bibinfo {author}
  {\bibfnamefont {V.}~\bibnamefont {Novosad}},\ }\href@noop {} {\bibfield
  {journal} {\bibinfo  {journal} {Journal of Applied Physics}\ }\textbf
  {\bibinfo {volume} {128}},\ \bibinfo {pages} {130902} (\bibinfo {year}
  {2020})}\BibitemShut {NoStop}%
\bibitem [{\citenamefont {Lachance-Quirion}\ \emph {et~al.}(2019)\citenamefont
  {Lachance-Quirion}, \citenamefont {Tabuchi}, \citenamefont {Gloppe},
  \citenamefont {Usami},\ and\ \citenamefont {Nakamura}}]{LachanceQuirion2019}%
  \BibitemOpen
  \bibfield  {author} {\bibinfo {author} {\bibfnamefont {D.}~\bibnamefont
  {Lachance-Quirion}}, \bibinfo {author} {\bibfnamefont {Y.}~\bibnamefont
  {Tabuchi}}, \bibinfo {author} {\bibfnamefont {A.}~\bibnamefont {Gloppe}},
  \bibinfo {author} {\bibfnamefont {K.}~\bibnamefont {Usami}},\ and\ \bibinfo
  {author} {\bibfnamefont {Y.}~\bibnamefont {Nakamura}},\ }\href@noop {}
  {\bibfield  {journal} {\bibinfo  {journal} {Applied Physics Express}\
  }\textbf {\bibinfo {volume} {12}},\ \bibinfo {pages} {070101} (\bibinfo
  {year} {2019})}\BibitemShut {NoStop}%
\bibitem [{\citenamefont {Tabuchi}\ \emph {et~al.}(2014)\citenamefont
  {Tabuchi}, \citenamefont {Ishino}, \citenamefont {Ishikawa}, \citenamefont
  {Yamazaki}, \citenamefont {Usami},\ and\ \citenamefont
  {Nakamura}}]{Tabuchi2014}%
  \BibitemOpen
  \bibfield  {author} {\bibinfo {author} {\bibfnamefont {Y.}~\bibnamefont
  {Tabuchi}}, \bibinfo {author} {\bibfnamefont {S.}~\bibnamefont {Ishino}},
  \bibinfo {author} {\bibfnamefont {T.}~\bibnamefont {Ishikawa}}, \bibinfo
  {author} {\bibfnamefont {R.}~\bibnamefont {Yamazaki}}, \bibinfo {author}
  {\bibfnamefont {K.}~\bibnamefont {Usami}},\ and\ \bibinfo {author}
  {\bibfnamefont {Y.}~\bibnamefont {Nakamura}},\ }\href@noop {} {\bibfield
  {journal} {\bibinfo  {journal} {Physical Review Letters}\ }\textbf {\bibinfo
  {volume} {113}},\ \bibinfo {pages} {083603} (\bibinfo {year}
  {2014})}\BibitemShut {NoStop}%
\bibitem [{\citenamefont {Huebl}\ \emph {et~al.}(2013)\citenamefont {Huebl},
  \citenamefont {Zollitsch}, \citenamefont {Lotze}, \citenamefont {Hocke},
  \citenamefont {Greifenstein}, \citenamefont {Marx}, \citenamefont {Gross},\
  and\ \citenamefont {Goennenwein}}]{Huebl2013}%
  \BibitemOpen
  \bibfield  {author} {\bibinfo {author} {\bibfnamefont {H.}~\bibnamefont
  {Huebl}}, \bibinfo {author} {\bibfnamefont {C.~W.}\ \bibnamefont
  {Zollitsch}}, \bibinfo {author} {\bibfnamefont {J.}~\bibnamefont {Lotze}},
  \bibinfo {author} {\bibfnamefont {F.}~\bibnamefont {Hocke}}, \bibinfo
  {author} {\bibfnamefont {M.}~\bibnamefont {Greifenstein}}, \bibinfo {author}
  {\bibfnamefont {A.}~\bibnamefont {Marx}}, \bibinfo {author} {\bibfnamefont
  {R.}~\bibnamefont {Gross}},\ and\ \bibinfo {author} {\bibfnamefont
  {S.~T.~B.}\ \bibnamefont {Goennenwein}},\ }\href@noop {} {\bibfield
  {journal} {\bibinfo  {journal} {Physical Review Letters}\ }\textbf {\bibinfo
  {volume} {111}},\ \bibinfo {pages} {127003} (\bibinfo {year}
  {2013})}\BibitemShut {NoStop}%
\bibitem [{\citenamefont {Golovchanskiy}\ \emph
  {et~al.}(2021{\natexlab{a}})\citenamefont {Golovchanskiy}, \citenamefont
  {Abramov}, \citenamefont {Stolyarov}, \citenamefont {Weides}, \citenamefont
  {Ryazanov}, \citenamefont {Golubov}, \citenamefont {Ustinov},\ and\
  \citenamefont {Kupriyanov}}]{Golovchanskiy2021}%
  \BibitemOpen
  \bibfield  {author} {\bibinfo {author} {\bibfnamefont {I.~A.}\ \bibnamefont
  {Golovchanskiy}}, \bibinfo {author} {\bibfnamefont {N.~N.}\ \bibnamefont
  {Abramov}}, \bibinfo {author} {\bibfnamefont {V.~S.}\ \bibnamefont
  {Stolyarov}}, \bibinfo {author} {\bibfnamefont {M.}~\bibnamefont {Weides}},
  \bibinfo {author} {\bibfnamefont {V.~V.}\ \bibnamefont {Ryazanov}}, \bibinfo
  {author} {\bibfnamefont {A.~A.}\ \bibnamefont {Golubov}}, \bibinfo {author}
  {\bibfnamefont {A.~V.}\ \bibnamefont {Ustinov}},\ and\ \bibinfo {author}
  {\bibfnamefont {M.~Y.}\ \bibnamefont {Kupriyanov}},\ }\href@noop {}
  {\bibfield  {journal} {\bibinfo  {journal} {Science Advances}\ }\textbf
  {\bibinfo {volume} {7}} (\bibinfo {year} {2021}{\natexlab{a}})}\BibitemShut
  {NoStop}%
\bibitem [{\citenamefont {Flower}\ \emph {et~al.}(2019)\citenamefont {Flower},
  \citenamefont {Goryachev}, \citenamefont {Bourhill},\ and\ \citenamefont
  {Tobar}}]{Flower2019}%
  \BibitemOpen
  \bibfield  {author} {\bibinfo {author} {\bibfnamefont {G.}~\bibnamefont
  {Flower}}, \bibinfo {author} {\bibfnamefont {M.}~\bibnamefont {Goryachev}},
  \bibinfo {author} {\bibfnamefont {J.}~\bibnamefont {Bourhill}},\ and\
  \bibinfo {author} {\bibfnamefont {M.~E.}\ \bibnamefont {Tobar}},\ }\href@noop
  {} {\bibfield  {journal} {\bibinfo  {journal} {New Journal of Physics}\
  }\textbf {\bibinfo {volume} {21}},\ \bibinfo {pages} {095004} (\bibinfo
  {year} {2019})}\BibitemShut {NoStop}%
\bibitem [{\citenamefont {Zhang}\ \emph {et~al.}(2014)\citenamefont {Zhang},
  \citenamefont {Zou}, \citenamefont {Jiang},\ and\ \citenamefont
  {Tang}}]{Zhang2014}%
  \BibitemOpen
  \bibfield  {author} {\bibinfo {author} {\bibfnamefont {X.}~\bibnamefont
  {Zhang}}, \bibinfo {author} {\bibfnamefont {C.-L.}\ \bibnamefont {Zou}},
  \bibinfo {author} {\bibfnamefont {L.}~\bibnamefont {Jiang}},\ and\ \bibinfo
  {author} {\bibfnamefont {H.~X.}\ \bibnamefont {Tang}},\ }\href@noop {}
  {\bibfield  {journal} {\bibinfo  {journal} {Physical Review Letters}\
  }\textbf {\bibinfo {volume} {113}},\ \bibinfo {pages} {156401} (\bibinfo
  {year} {2014})}\BibitemShut {NoStop}%
\bibitem [{\citenamefont {Golovchanskiy}\ \emph
  {et~al.}(2021{\natexlab{b}})\citenamefont {Golovchanskiy}, \citenamefont
  {Abramov}, \citenamefont {Stolyarov}, \citenamefont {Golubov}, \citenamefont
  {Kupriyanov}, \citenamefont {Ryazanov},\ and\ \citenamefont
  {Ustinov}}]{Golovchanskiy2021a}%
  \BibitemOpen
  \bibfield  {author} {\bibinfo {author} {\bibfnamefont {I.}~\bibnamefont
  {Golovchanskiy}}, \bibinfo {author} {\bibfnamefont {N.}~\bibnamefont
  {Abramov}}, \bibinfo {author} {\bibfnamefont {V.}~\bibnamefont {Stolyarov}},
  \bibinfo {author} {\bibfnamefont {A.}~\bibnamefont {Golubov}}, \bibinfo
  {author} {\bibfnamefont {M.~Y.}\ \bibnamefont {Kupriyanov}}, \bibinfo
  {author} {\bibfnamefont {V.}~\bibnamefont {Ryazanov}},\ and\ \bibinfo
  {author} {\bibfnamefont {A.}~\bibnamefont {Ustinov}},\ }\href@noop {}
  {\bibfield  {journal} {\bibinfo  {journal} {Physical Review Applied}\
  }\textbf {\bibinfo {volume} {16}},\ \bibinfo {pages} {034029} (\bibinfo
  {year} {2021}{\natexlab{b}})}\BibitemShut {NoStop}%
\bibitem [{\citenamefont {Lambert}\ \emph {et~al.}(2020)\citenamefont
  {Lambert}, \citenamefont {Rueda}, \citenamefont {Sedlmeir},\ and\
  \citenamefont {Schwefel}}]{Lambert2020}%
  \BibitemOpen
  \bibfield  {author} {\bibinfo {author} {\bibfnamefont {N.~J.}\ \bibnamefont
  {Lambert}}, \bibinfo {author} {\bibfnamefont {A.}~\bibnamefont {Rueda}},
  \bibinfo {author} {\bibfnamefont {F.}~\bibnamefont {Sedlmeir}},\ and\
  \bibinfo {author} {\bibfnamefont {H.~G.~L.}\ \bibnamefont {Schwefel}},\
  }\href@noop {} {\bibfield  {journal} {\bibinfo  {journal} {Advanced Quantum
  Technologies}\ }\textbf {\bibinfo {volume} {3}},\ \bibinfo {pages} {1900077}
  (\bibinfo {year} {2020})}\BibitemShut {NoStop}%
\bibitem [{\citenamefont {Hisatomi}\ \emph {et~al.}(2016)\citenamefont
  {Hisatomi}, \citenamefont {Osada}, \citenamefont {Tabuchi}, \citenamefont
  {Ishikawa}, \citenamefont {Noguchi}, \citenamefont {Yamazaki}, \citenamefont
  {Usami},\ and\ \citenamefont {Nakamura}}]{Hisatomi2016}%
  \BibitemOpen
  \bibfield  {author} {\bibinfo {author} {\bibfnamefont {R.}~\bibnamefont
  {Hisatomi}}, \bibinfo {author} {\bibfnamefont {A.}~\bibnamefont {Osada}},
  \bibinfo {author} {\bibfnamefont {Y.}~\bibnamefont {Tabuchi}}, \bibinfo
  {author} {\bibfnamefont {T.}~\bibnamefont {Ishikawa}}, \bibinfo {author}
  {\bibfnamefont {A.}~\bibnamefont {Noguchi}}, \bibinfo {author} {\bibfnamefont
  {R.}~\bibnamefont {Yamazaki}}, \bibinfo {author} {\bibfnamefont
  {K.}~\bibnamefont {Usami}},\ and\ \bibinfo {author} {\bibfnamefont
  {Y.}~\bibnamefont {Nakamura}},\ }\href@noop {} {\bibfield  {journal}
  {\bibinfo  {journal} {Physical Review B}\ }\textbf {\bibinfo {volume} {93}},\
  \bibinfo {pages} {174427} (\bibinfo {year} {2016})}\BibitemShut {NoStop}%
\bibitem [{\citenamefont {Li}\ \emph {et~al.}(2022)\citenamefont {Li},
  \citenamefont {Yefremenko}, \citenamefont {Lisovenko}, \citenamefont
  {Trevillian}, \citenamefont {Polakovic}, \citenamefont {Cecil}, \citenamefont
  {Barry}, \citenamefont {Pearson}, \citenamefont {Divan}, \citenamefont
  {Tyberkevych}, \citenamefont {Chang}, \citenamefont {Welp}, \citenamefont
  {Kwok},\ and\ \citenamefont {Novosad}}]{Li2022}%
  \BibitemOpen
  \bibfield  {author} {\bibinfo {author} {\bibfnamefont {Y.}~\bibnamefont
  {Li}}, \bibinfo {author} {\bibfnamefont {V.~G.}\ \bibnamefont {Yefremenko}},
  \bibinfo {author} {\bibfnamefont {M.}~\bibnamefont {Lisovenko}}, \bibinfo
  {author} {\bibfnamefont {C.}~\bibnamefont {Trevillian}}, \bibinfo {author}
  {\bibfnamefont {T.}~\bibnamefont {Polakovic}}, \bibinfo {author}
  {\bibfnamefont {T.~W.}\ \bibnamefont {Cecil}}, \bibinfo {author}
  {\bibfnamefont {P.~S.}\ \bibnamefont {Barry}}, \bibinfo {author}
  {\bibfnamefont {J.}~\bibnamefont {Pearson}}, \bibinfo {author} {\bibfnamefont
  {R.}~\bibnamefont {Divan}}, \bibinfo {author} {\bibfnamefont
  {V.}~\bibnamefont {Tyberkevych}}, \bibinfo {author} {\bibfnamefont {C.~L.}\
  \bibnamefont {Chang}}, \bibinfo {author} {\bibfnamefont {U.}~\bibnamefont
  {Welp}}, \bibinfo {author} {\bibfnamefont {W.-K.}\ \bibnamefont {Kwok}},\
  and\ \bibinfo {author} {\bibfnamefont {V.}~\bibnamefont {Novosad}},\
  }\href@noop {} {\bibfield  {journal} {\bibinfo  {journal} {Physical Review
  Letters}\ }\textbf {\bibinfo {volume} {128}},\ \bibinfo {pages} {047701}
  (\bibinfo {year} {2022})}\BibitemShut {NoStop}%
\bibitem [{\citenamefont {Lachance-Quirion}\ \emph {et~al.}(2020)\citenamefont
  {Lachance-Quirion}, \citenamefont {Wolski}, \citenamefont {Tabuchi},
  \citenamefont {Kono}, \citenamefont {Usami},\ and\ \citenamefont
  {Nakamura}}]{LachanceQuirion2020}%
  \BibitemOpen
  \bibfield  {author} {\bibinfo {author} {\bibfnamefont {D.}~\bibnamefont
  {Lachance-Quirion}}, \bibinfo {author} {\bibfnamefont {S.~P.}\ \bibnamefont
  {Wolski}}, \bibinfo {author} {\bibfnamefont {Y.}~\bibnamefont {Tabuchi}},
  \bibinfo {author} {\bibfnamefont {S.}~\bibnamefont {Kono}}, \bibinfo {author}
  {\bibfnamefont {K.}~\bibnamefont {Usami}},\ and\ \bibinfo {author}
  {\bibfnamefont {Y.}~\bibnamefont {Nakamura}},\ }\href@noop {} {\bibfield
  {journal} {\bibinfo  {journal} {Science}\ }\textbf {\bibinfo {volume}
  {367}},\ \bibinfo {pages} {425} (\bibinfo {year} {2020})}\BibitemShut
  {NoStop}%
\bibitem [{\citenamefont {Wolski}\ \emph {et~al.}(2020)\citenamefont {Wolski},
  \citenamefont {Lachance-Quirion}, \citenamefont {Tabuchi}, \citenamefont
  {Kono}, \citenamefont {Noguchi}, \citenamefont {Usami},\ and\ \citenamefont
  {Nakamura}}]{Wolski2020}%
  \BibitemOpen
  \bibfield  {author} {\bibinfo {author} {\bibfnamefont {S.}~\bibnamefont
  {Wolski}}, \bibinfo {author} {\bibfnamefont {D.}~\bibnamefont
  {Lachance-Quirion}}, \bibinfo {author} {\bibfnamefont {Y.}~\bibnamefont
  {Tabuchi}}, \bibinfo {author} {\bibfnamefont {S.}~\bibnamefont {Kono}},
  \bibinfo {author} {\bibfnamefont {A.}~\bibnamefont {Noguchi}}, \bibinfo
  {author} {\bibfnamefont {K.}~\bibnamefont {Usami}},\ and\ \bibinfo {author}
  {\bibfnamefont {Y.}~\bibnamefont {Nakamura}},\ }\href@noop {} {\bibfield
  {journal} {\bibinfo  {journal} {Physical Review Letters}\ }\textbf {\bibinfo
  {volume} {125}},\ \bibinfo {pages} {117701} (\bibinfo {year}
  {2020})}\BibitemShut {NoStop}%
\bibitem [{\citenamefont {Wang}\ \emph {et~al.}(2021)\citenamefont {Wang},
  \citenamefont {van Geldern}, \citenamefont {Connolly}, \citenamefont {Wang},
  \citenamefont {Shilcusky}, \citenamefont {McDonald}, \citenamefont {Clerk},\
  and\ \citenamefont {Wang}}]{Wang2021}%
  \BibitemOpen
  \bibfield  {author} {\bibinfo {author} {\bibfnamefont {Y.-Y.}\ \bibnamefont
  {Wang}}, \bibinfo {author} {\bibfnamefont {S.}~\bibnamefont {van Geldern}},
  \bibinfo {author} {\bibfnamefont {T.}~\bibnamefont {Connolly}}, \bibinfo
  {author} {\bibfnamefont {Y.-X.}\ \bibnamefont {Wang}}, \bibinfo {author}
  {\bibfnamefont {A.}~\bibnamefont {Shilcusky}}, \bibinfo {author}
  {\bibfnamefont {A.}~\bibnamefont {McDonald}}, \bibinfo {author}
  {\bibfnamefont {A.~A.}\ \bibnamefont {Clerk}},\ and\ \bibinfo {author}
  {\bibfnamefont {C.}~\bibnamefont {Wang}},\ }\href@noop {} {\bibfield
  {journal} {\bibinfo  {journal} {Physical Review Applied}\ }\textbf {\bibinfo
  {volume} {16}},\ \bibinfo {pages} {064066} (\bibinfo {year}
  {2021})}\BibitemShut {NoStop}%
\bibitem [{\citenamefont {Zhu}\ \emph {et~al.}(2020)\citenamefont {Zhu},
  \citenamefont {Han}, \citenamefont {Zou}, \citenamefont {Xu},\ and\
  \citenamefont {Tang}}]{Zhu2020}%
  \BibitemOpen
  \bibfield  {author} {\bibinfo {author} {\bibfnamefont {N.}~\bibnamefont
  {Zhu}}, \bibinfo {author} {\bibfnamefont {X.}~\bibnamefont {Han}}, \bibinfo
  {author} {\bibfnamefont {C.-L.}\ \bibnamefont {Zou}}, \bibinfo {author}
  {\bibfnamefont {M.}~\bibnamefont {Xu}},\ and\ \bibinfo {author}
  {\bibfnamefont {H.~X.}\ \bibnamefont {Tang}},\ }\href@noop {} {\bibfield
  {journal} {\bibinfo  {journal} {Physical Review A}\ }\textbf {\bibinfo
  {volume} {101}},\ \bibinfo {pages} {043842} (\bibinfo {year}
  {2020})}\BibitemShut {NoStop}%
\bibitem [{\citenamefont {Zhang}\ \emph {et~al.}(2020)\citenamefont {Zhang},
  \citenamefont {Galda}, \citenamefont {Han}, \citenamefont {Jin},\ and\
  \citenamefont {Vinokur}}]{Zhang2020}%
  \BibitemOpen
  \bibfield  {author} {\bibinfo {author} {\bibfnamefont {X.}~\bibnamefont
  {Zhang}}, \bibinfo {author} {\bibfnamefont {A.}~\bibnamefont {Galda}},
  \bibinfo {author} {\bibfnamefont {X.}~\bibnamefont {Han}}, \bibinfo {author}
  {\bibfnamefont {D.}~\bibnamefont {Jin}},\ and\ \bibinfo {author}
  {\bibfnamefont {V.~M.}\ \bibnamefont {Vinokur}},\ }\href@noop {} {\bibfield
  {journal} {\bibinfo  {journal} {Physical Review Applied}\ }\textbf {\bibinfo
  {volume} {13}},\ \bibinfo {pages} {044039} (\bibinfo {year}
  {2020})}\BibitemShut {NoStop}%
\bibitem [{\citenamefont {Wang}\ \emph {et~al.}(2019)\citenamefont {Wang},
  \citenamefont {Rao}, \citenamefont {Yang}, \citenamefont {Xu}, \citenamefont
  {Gui}, \citenamefont {Yao}, \citenamefont {You},\ and\ \citenamefont
  {Hu}}]{Wang2019}%
  \BibitemOpen
  \bibfield  {author} {\bibinfo {author} {\bibfnamefont {Y.-P.}\ \bibnamefont
  {Wang}}, \bibinfo {author} {\bibfnamefont {J.}~\bibnamefont {Rao}}, \bibinfo
  {author} {\bibfnamefont {Y.}~\bibnamefont {Yang}}, \bibinfo {author}
  {\bibfnamefont {P.-C.}\ \bibnamefont {Xu}}, \bibinfo {author} {\bibfnamefont
  {Y.}~\bibnamefont {Gui}}, \bibinfo {author} {\bibfnamefont {B.}~\bibnamefont
  {Yao}}, \bibinfo {author} {\bibfnamefont {J.}~\bibnamefont {You}},\ and\
  \bibinfo {author} {\bibfnamefont {C.-M.}\ \bibnamefont {Hu}},\ }\href@noop {}
  {\bibfield  {journal} {\bibinfo  {journal} {Physical Review Letters}\
  }\textbf {\bibinfo {volume} {123}},\ \bibinfo {pages} {127202} (\bibinfo
  {year} {2019})}\BibitemShut {NoStop}%
\bibitem [{\citenamefont {Pozar}(2011)}]{Pozar2011}%
  \BibitemOpen
  \bibfield  {author} {\bibinfo {author} {\bibfnamefont {D.~M.}\ \bibnamefont
  {Pozar}},\ }\href@noop {} {\emph {\bibinfo {title} {Microwave engineering}}}\
  (\bibinfo  {publisher} {Wiley},\ \bibinfo {year} {2011})\BibitemShut
  {NoStop}%
\bibitem [{\citenamefont {Kittel}\ and\ \citenamefont
  {Holcomb}(1967)}]{Kittel1967}%
  \BibitemOpen
  \bibfield  {author} {\bibinfo {author} {\bibfnamefont {C.}~\bibnamefont
  {Kittel}}\ and\ \bibinfo {author} {\bibfnamefont {D.~F.}\ \bibnamefont
  {Holcomb}},\ }\href@noop {} {\bibfield  {journal} {\bibinfo  {journal}
  {American Journal of Physics}\ }\textbf {\bibinfo {volume} {35}},\ \bibinfo
  {pages} {547} (\bibinfo {year} {1967})}\BibitemShut {NoStop}%
\bibitem [{\citenamefont {Walker}(1957)}]{Walker1957}%
  \BibitemOpen
  \bibfield  {author} {\bibinfo {author} {\bibfnamefont {L.~R.}\ \bibnamefont
  {Walker}},\ }\href@noop {} {\bibfield  {journal} {\bibinfo  {journal}
  {Physical Review}\ }\textbf {\bibinfo {volume} {105}},\ \bibinfo {pages}
  {390} (\bibinfo {year} {1957})}\BibitemShut {NoStop}%
\bibitem [{\citenamefont {Baity}\ \emph {et~al.}(2021)\citenamefont {Baity},
  \citenamefont {Bozhko}, \citenamefont {Mac{\^{e}}do}, \citenamefont {Smith},
  \citenamefont {Holland}, \citenamefont {Danilin}, \citenamefont {Seferai},
  \citenamefont {Barbosa}, \citenamefont {Peroor}, \citenamefont {Goldman},
  \citenamefont {Nasti}, \citenamefont {Paul}, \citenamefont {Hadfield},
  \citenamefont {McVitie},\ and\ \citenamefont {Weides}}]{Baity2021}%
  \BibitemOpen
  \bibfield  {author} {\bibinfo {author} {\bibfnamefont {P.~G.}\ \bibnamefont
  {Baity}}, \bibinfo {author} {\bibfnamefont {D.~A.}\ \bibnamefont {Bozhko}},
  \bibinfo {author} {\bibfnamefont {R.}~\bibnamefont {Mac{\^{e}}do}}, \bibinfo
  {author} {\bibfnamefont {W.}~\bibnamefont {Smith}}, \bibinfo {author}
  {\bibfnamefont {R.~C.}\ \bibnamefont {Holland}}, \bibinfo {author}
  {\bibfnamefont {S.}~\bibnamefont {Danilin}}, \bibinfo {author} {\bibfnamefont
  {V.}~\bibnamefont {Seferai}}, \bibinfo {author} {\bibfnamefont
  {J.}~\bibnamefont {Barbosa}}, \bibinfo {author} {\bibfnamefont {R.~R.}\
  \bibnamefont {Peroor}}, \bibinfo {author} {\bibfnamefont {S.}~\bibnamefont
  {Goldman}}, \bibinfo {author} {\bibfnamefont {U.}~\bibnamefont {Nasti}},
  \bibinfo {author} {\bibfnamefont {J.}~\bibnamefont {Paul}}, \bibinfo {author}
  {\bibfnamefont {R.~H.}\ \bibnamefont {Hadfield}}, \bibinfo {author}
  {\bibfnamefont {S.}~\bibnamefont {McVitie}},\ and\ \bibinfo {author}
  {\bibfnamefont {M.}~\bibnamefont {Weides}},\ }\href@noop {} {\bibfield
  {journal} {\bibinfo  {journal} {Applied Physics Letters}\ }\textbf {\bibinfo
  {volume} {119}},\ \bibinfo {pages} {033502} (\bibinfo {year}
  {2021})}\BibitemShut {NoStop}%
\bibitem [{\citenamefont {Mihalceanu}\ \emph {et~al.}(2018)\citenamefont
  {Mihalceanu}, \citenamefont {Vasyuchka}, \citenamefont {Bozhko},
  \citenamefont {Langner}, \citenamefont {Nechiporuk}, \citenamefont
  {Romanyuk}, \citenamefont {Hillebrands},\ and\ \citenamefont
  {Serga}}]{Mihalceanu2018}%
  \BibitemOpen
  \bibfield  {author} {\bibinfo {author} {\bibfnamefont {L.}~\bibnamefont
  {Mihalceanu}}, \bibinfo {author} {\bibfnamefont {V.~I.}\ \bibnamefont
  {Vasyuchka}}, \bibinfo {author} {\bibfnamefont {D.~A.}\ \bibnamefont
  {Bozhko}}, \bibinfo {author} {\bibfnamefont {T.}~\bibnamefont {Langner}},
  \bibinfo {author} {\bibfnamefont {A.~Y.}\ \bibnamefont {Nechiporuk}},
  \bibinfo {author} {\bibfnamefont {V.~F.}\ \bibnamefont {Romanyuk}}, \bibinfo
  {author} {\bibfnamefont {B.}~\bibnamefont {Hillebrands}},\ and\ \bibinfo
  {author} {\bibfnamefont {A.~A.}\ \bibnamefont {Serga}},\ }\href@noop {}
  {\bibfield  {journal} {\bibinfo  {journal} {Physical Review B}\ }\textbf
  {\bibinfo {volume} {97}},\ \bibinfo {pages} {214405} (\bibinfo {year}
  {2018})}\BibitemShut {NoStop}%
\bibitem [{\citenamefont {Hou}\ and\ \citenamefont {Liu}(2019)}]{Hou2019}%
  \BibitemOpen
  \bibfield  {author} {\bibinfo {author} {\bibfnamefont {J.~T.}\ \bibnamefont
  {Hou}}\ and\ \bibinfo {author} {\bibfnamefont {L.}~\bibnamefont {Liu}},\
  }\href@noop {} {\bibfield  {journal} {\bibinfo  {journal} {Physical Review
  Letters}\ }\textbf {\bibinfo {volume} {123}},\ \bibinfo {pages} {107702}
  (\bibinfo {year} {2019})}\BibitemShut {NoStop}%
\bibitem [{\citenamefont {Li}\ \emph {et~al.}(2019)\citenamefont {Li},
  \citenamefont {Polakovic}, \citenamefont {Wang}, \citenamefont {Xu},
  \citenamefont {Lendinez}, \citenamefont {Zhang}, \citenamefont {Ding},
  \citenamefont {Khaire}, \citenamefont {Saglam}, \citenamefont {Divan},
  \citenamefont {Pearson}, \citenamefont {Kwok}, \citenamefont {Xiao},
  \citenamefont {Novosad}, \citenamefont {Hoffmann},\ and\ \citenamefont
  {Zhang}}]{Li2019}%
  \BibitemOpen
  \bibfield  {author} {\bibinfo {author} {\bibfnamefont {Y.}~\bibnamefont
  {Li}}, \bibinfo {author} {\bibfnamefont {T.}~\bibnamefont {Polakovic}},
  \bibinfo {author} {\bibfnamefont {Y.-L.}\ \bibnamefont {Wang}}, \bibinfo
  {author} {\bibfnamefont {J.}~\bibnamefont {Xu}}, \bibinfo {author}
  {\bibfnamefont {S.}~\bibnamefont {Lendinez}}, \bibinfo {author}
  {\bibfnamefont {Z.}~\bibnamefont {Zhang}}, \bibinfo {author} {\bibfnamefont
  {J.}~\bibnamefont {Ding}}, \bibinfo {author} {\bibfnamefont {T.}~\bibnamefont
  {Khaire}}, \bibinfo {author} {\bibfnamefont {H.}~\bibnamefont {Saglam}},
  \bibinfo {author} {\bibfnamefont {R.}~\bibnamefont {Divan}}, \bibinfo
  {author} {\bibfnamefont {J.}~\bibnamefont {Pearson}}, \bibinfo {author}
  {\bibfnamefont {W.-K.}\ \bibnamefont {Kwok}}, \bibinfo {author}
  {\bibfnamefont {Z.}~\bibnamefont {Xiao}}, \bibinfo {author} {\bibfnamefont
  {V.}~\bibnamefont {Novosad}}, \bibinfo {author} {\bibfnamefont
  {A.}~\bibnamefont {Hoffmann}},\ and\ \bibinfo {author} {\bibfnamefont
  {W.}~\bibnamefont {Zhang}},\ }\href@noop {} {\bibfield  {journal} {\bibinfo
  {journal} {Physical Review Letters}\ }\textbf {\bibinfo {volume} {123}},\
  \bibinfo {pages} {107701} (\bibinfo {year} {2019})}\BibitemShut {NoStop}%
\bibitem [{\citenamefont {Golovchanskiy}\ \emph {et~al.}(2018)\citenamefont
  {Golovchanskiy}, \citenamefont {Abramov}, \citenamefont {Stolyarov},
  \citenamefont {Shchetinin}, \citenamefont {Dzhumaev}, \citenamefont
  {Averkin}, \citenamefont {Kozlov}, \citenamefont {Golubov}, \citenamefont
  {Ryazanov},\ and\ \citenamefont {Ustinov}}]{Golovchanskiy2018}%
  \BibitemOpen
  \bibfield  {author} {\bibinfo {author} {\bibfnamefont {I.~A.}\ \bibnamefont
  {Golovchanskiy}}, \bibinfo {author} {\bibfnamefont {N.~N.}\ \bibnamefont
  {Abramov}}, \bibinfo {author} {\bibfnamefont {V.~S.}\ \bibnamefont
  {Stolyarov}}, \bibinfo {author} {\bibfnamefont {I.~V.}\ \bibnamefont
  {Shchetinin}}, \bibinfo {author} {\bibfnamefont {P.~S.}\ \bibnamefont
  {Dzhumaev}}, \bibinfo {author} {\bibfnamefont {A.~S.}\ \bibnamefont
  {Averkin}}, \bibinfo {author} {\bibfnamefont {S.~N.}\ \bibnamefont {Kozlov}},
  \bibinfo {author} {\bibfnamefont {A.~A.}\ \bibnamefont {Golubov}}, \bibinfo
  {author} {\bibfnamefont {V.~V.}\ \bibnamefont {Ryazanov}},\ and\ \bibinfo
  {author} {\bibfnamefont {A.~V.}\ \bibnamefont {Ustinov}},\ }\href@noop {}
  {\bibfield  {journal} {\bibinfo  {journal} {Journal of Applied Physics}\
  }\textbf {\bibinfo {volume} {123}},\ \bibinfo {pages} {173904} (\bibinfo
  {year} {2018})}\BibitemShut {NoStop}%
\bibitem [{\citenamefont {Haygood}\ \emph {et~al.}(2021)\citenamefont
  {Haygood}, \citenamefont {Pufall}, \citenamefont {Edwards}, \citenamefont
  {Shaw},\ and\ \citenamefont {Rippard}}]{Haygood2021}%
  \BibitemOpen
  \bibfield  {author} {\bibinfo {author} {\bibfnamefont {I.}~\bibnamefont
  {Haygood}}, \bibinfo {author} {\bibfnamefont {M.}~\bibnamefont {Pufall}},
  \bibinfo {author} {\bibfnamefont {E.}~\bibnamefont {Edwards}}, \bibinfo
  {author} {\bibfnamefont {J.~M.}\ \bibnamefont {Shaw}},\ and\ \bibinfo
  {author} {\bibfnamefont {W.}~\bibnamefont {Rippard}},\ }\href@noop {}
  {\bibfield  {journal} {\bibinfo  {journal} {Physical Review Applied}\
  }\textbf {\bibinfo {volume} {15}},\ \bibinfo {pages} {054021} (\bibinfo
  {year} {2021})}\BibitemShut {NoStop}%
\bibitem [{\citenamefont {Schoen}\ \emph {et~al.}(2016)\citenamefont {Schoen},
  \citenamefont {Thonig}, \citenamefont {Schneider}, \citenamefont {Silva},
  \citenamefont {Nembach}, \citenamefont {Eriksson}, \citenamefont {Karis},\
  and\ \citenamefont {Shaw}}]{Schoen2016}%
  \BibitemOpen
  \bibfield  {author} {\bibinfo {author} {\bibfnamefont {M.~A.~W.}\
  \bibnamefont {Schoen}}, \bibinfo {author} {\bibfnamefont {D.}~\bibnamefont
  {Thonig}}, \bibinfo {author} {\bibfnamefont {M.~L.}\ \bibnamefont
  {Schneider}}, \bibinfo {author} {\bibfnamefont {T.~J.}\ \bibnamefont
  {Silva}}, \bibinfo {author} {\bibfnamefont {H.~T.}\ \bibnamefont {Nembach}},
  \bibinfo {author} {\bibfnamefont {O.}~\bibnamefont {Eriksson}}, \bibinfo
  {author} {\bibfnamefont {O.}~\bibnamefont {Karis}},\ and\ \bibinfo {author}
  {\bibfnamefont {J.~M.}\ \bibnamefont {Shaw}},\ }\href@noop {} {\bibfield
  {journal} {\bibinfo  {journal} {Nature Physics}\ }\textbf {\bibinfo {volume}
  {12}},\ \bibinfo {pages} {839} (\bibinfo {year} {2016})}\BibitemShut
  {NoStop}%
\bibitem [{\citenamefont {Gimeno}\ \emph {et~al.}(2020)\citenamefont {Gimeno},
  \citenamefont {Kersten}, \citenamefont {Pallar{\'{e}}s}, \citenamefont
  {Hermosilla}, \citenamefont {Mart{\'{\i}}nez-P{\'{e}}rez}, \citenamefont
  {Jenkins}, \citenamefont {Angerer}, \citenamefont {S{\'{a}}nchez-Azqueta},
  \citenamefont {Zueco}, \citenamefont {Majer}, \citenamefont {Lostao},\ and\
  \citenamefont {Luis}}]{Gimeno2020}%
  \BibitemOpen
  \bibfield  {author} {\bibinfo {author} {\bibfnamefont {I.}~\bibnamefont
  {Gimeno}}, \bibinfo {author} {\bibfnamefont {W.}~\bibnamefont {Kersten}},
  \bibinfo {author} {\bibfnamefont {M.~C.}\ \bibnamefont {Pallar{\'{e}}s}},
  \bibinfo {author} {\bibfnamefont {P.}~\bibnamefont {Hermosilla}}, \bibinfo
  {author} {\bibfnamefont {M.~J.}\ \bibnamefont {Mart{\'{\i}}nez-P{\'{e}}rez}},
  \bibinfo {author} {\bibfnamefont {M.~D.}\ \bibnamefont {Jenkins}}, \bibinfo
  {author} {\bibfnamefont {A.}~\bibnamefont {Angerer}}, \bibinfo {author}
  {\bibfnamefont {C.}~\bibnamefont {S{\'{a}}nchez-Azqueta}}, \bibinfo {author}
  {\bibfnamefont {D.}~\bibnamefont {Zueco}}, \bibinfo {author} {\bibfnamefont
  {J.}~\bibnamefont {Majer}}, \bibinfo {author} {\bibfnamefont
  {A.}~\bibnamefont {Lostao}},\ and\ \bibinfo {author} {\bibfnamefont
  {F.}~\bibnamefont {Luis}},\ }\href@noop {} {\bibfield  {journal} {\bibinfo
  {journal} {{ACS} Nano}\ }\textbf {\bibinfo {volume} {14}},\ \bibinfo {pages}
  {8707} (\bibinfo {year} {2020})}\BibitemShut {NoStop}%
\bibitem [{\citenamefont {Haikka}\ \emph {et~al.}(2017)\citenamefont {Haikka},
  \citenamefont {Kubo}, \citenamefont {Bienfait}, \citenamefont {Bertet},\ and\
  \citenamefont {M{\o}lmer}}]{Haikka2017}%
  \BibitemOpen
  \bibfield  {author} {\bibinfo {author} {\bibfnamefont {P.}~\bibnamefont
  {Haikka}}, \bibinfo {author} {\bibfnamefont {Y.}~\bibnamefont {Kubo}},
  \bibinfo {author} {\bibfnamefont {A.}~\bibnamefont {Bienfait}}, \bibinfo
  {author} {\bibfnamefont {P.}~\bibnamefont {Bertet}},\ and\ \bibinfo {author}
  {\bibfnamefont {K.}~\bibnamefont {M{\o}lmer}},\ }\href@noop {} {\bibfield
  {journal} {\bibinfo  {journal} {Physical Review A}\ }\textbf {\bibinfo
  {volume} {95}},\ \bibinfo {pages} {022306} (\bibinfo {year}
  {2017})}\BibitemShut {NoStop}%
\bibitem [{\citenamefont {Jenkins}\ \emph {et~al.}(2014)\citenamefont
  {Jenkins}, \citenamefont {Naether}, \citenamefont {Ciria}, \citenamefont
  {Ses{\'{e}}}, \citenamefont {Atkinson}, \citenamefont
  {S{\'{a}}nchez-Azqueta}, \citenamefont {del Barco}, \citenamefont {Majer},
  \citenamefont {Zueco},\ and\ \citenamefont {Luis}}]{Jenkins2014}%
  \BibitemOpen
  \bibfield  {author} {\bibinfo {author} {\bibfnamefont {M.~D.}\ \bibnamefont
  {Jenkins}}, \bibinfo {author} {\bibfnamefont {U.}~\bibnamefont {Naether}},
  \bibinfo {author} {\bibfnamefont {M.}~\bibnamefont {Ciria}}, \bibinfo
  {author} {\bibfnamefont {J.}~\bibnamefont {Ses{\'{e}}}}, \bibinfo {author}
  {\bibfnamefont {J.}~\bibnamefont {Atkinson}}, \bibinfo {author}
  {\bibfnamefont {C.}~\bibnamefont {S{\'{a}}nchez-Azqueta}}, \bibinfo {author}
  {\bibfnamefont {E.}~\bibnamefont {del Barco}}, \bibinfo {author}
  {\bibfnamefont {J.}~\bibnamefont {Majer}}, \bibinfo {author} {\bibfnamefont
  {D.}~\bibnamefont {Zueco}},\ and\ \bibinfo {author} {\bibfnamefont
  {F.}~\bibnamefont {Luis}},\ }\href@noop {} {\bibfield  {journal} {\bibinfo
  {journal} {Applied Physics Letters}\ }\textbf {\bibinfo {volume} {105}},\
  \bibinfo {pages} {162601} (\bibinfo {year} {2014})}\BibitemShut {NoStop}%
\bibitem [{\citenamefont {Probst}\ \emph {et~al.}(2017)\citenamefont {Probst},
  \citenamefont {Bienfait}, \citenamefont {Campagne-Ibarcq}, \citenamefont
  {Pla}, \citenamefont {Albanese}, \citenamefont {Barbosa}, \citenamefont
  {Schenkel}, \citenamefont {Vion}, \citenamefont {Esteve}, \citenamefont
  {M{\o}lmer}, \citenamefont {Morton}, \citenamefont {Heeres},\ and\
  \citenamefont {Bertet}}]{Probst2017}%
  \BibitemOpen
  \bibfield  {author} {\bibinfo {author} {\bibfnamefont {S.}~\bibnamefont
  {Probst}}, \bibinfo {author} {\bibfnamefont {A.}~\bibnamefont {Bienfait}},
  \bibinfo {author} {\bibfnamefont {P.}~\bibnamefont {Campagne-Ibarcq}},
  \bibinfo {author} {\bibfnamefont {J.~J.}\ \bibnamefont {Pla}}, \bibinfo
  {author} {\bibfnamefont {B.}~\bibnamefont {Albanese}}, \bibinfo {author}
  {\bibfnamefont {J.~F. D.~S.}\ \bibnamefont {Barbosa}}, \bibinfo {author}
  {\bibfnamefont {T.}~\bibnamefont {Schenkel}}, \bibinfo {author}
  {\bibfnamefont {D.}~\bibnamefont {Vion}}, \bibinfo {author} {\bibfnamefont
  {D.}~\bibnamefont {Esteve}}, \bibinfo {author} {\bibfnamefont
  {K.}~\bibnamefont {M{\o}lmer}}, \bibinfo {author} {\bibfnamefont {J.~J.~L.}\
  \bibnamefont {Morton}}, \bibinfo {author} {\bibfnamefont {R.}~\bibnamefont
  {Heeres}},\ and\ \bibinfo {author} {\bibfnamefont {P.}~\bibnamefont
  {Bertet}},\ }\href@noop {} {\bibfield  {journal} {\bibinfo  {journal}
  {Applied Physics Letters}\ }\textbf {\bibinfo {volume} {111}},\ \bibinfo
  {pages} {202604} (\bibinfo {year} {2017})}\BibitemShut {NoStop}%
\bibitem [{\citenamefont {Eichler}\ \emph {et~al.}(2017)\citenamefont
  {Eichler}, \citenamefont {Sigillito}, \citenamefont {Lyon},\ and\
  \citenamefont {Petta}}]{Eichler2017}%
  \BibitemOpen
  \bibfield  {author} {\bibinfo {author} {\bibfnamefont {C.}~\bibnamefont
  {Eichler}}, \bibinfo {author} {\bibfnamefont {A.}~\bibnamefont {Sigillito}},
  \bibinfo {author} {\bibfnamefont {S.}~\bibnamefont {Lyon}},\ and\ \bibinfo
  {author} {\bibfnamefont {J.}~\bibnamefont {Petta}},\ }\href@noop {}
  {\bibfield  {journal} {\bibinfo  {journal} {Physical Review Letters}\
  }\textbf {\bibinfo {volume} {118}},\ \bibinfo {pages} {037701} (\bibinfo
  {year} {2017})}\BibitemShut {NoStop}%
\bibitem [{\citenamefont {Mart{\'{i}}nez-P{\'{e}}rez}\ \emph
  {et~al.}(2018)\citenamefont {Mart{\'{i}}nez-P{\'{e}}rez}, \citenamefont
  {Pablo-Navarro}, \citenamefont {Müller}, \citenamefont {Kleiner},
  \citenamefont {Mag{\'{e}}n}, \citenamefont {Koelle}, \citenamefont
  {de~Teresa},\ and\ \citenamefont {Sese}}]{MartinezPerez2018a}%
  \BibitemOpen
  \bibfield  {author} {\bibinfo {author} {\bibfnamefont {M.~J.}\ \bibnamefont
  {Mart{\'{i}}nez-P{\'{e}}rez}}, \bibinfo {author} {\bibfnamefont
  {J.}~\bibnamefont {Pablo-Navarro}}, \bibinfo {author} {\bibfnamefont
  {B.}~\bibnamefont {Müller}}, \bibinfo {author} {\bibfnamefont
  {R.}~\bibnamefont {Kleiner}}, \bibinfo {author} {\bibfnamefont
  {C.}~\bibnamefont {Mag{\'{e}}n}}, \bibinfo {author} {\bibfnamefont
  {D.}~\bibnamefont {Koelle}}, \bibinfo {author} {\bibfnamefont {J.~M.}\
  \bibnamefont {de~Teresa}},\ and\ \bibinfo {author} {\bibfnamefont
  {J.}~\bibnamefont {Sese}},\ }\href@noop {} {\bibfield  {journal} {\bibinfo
  {journal} {Nano Letters}\ }\textbf {\bibinfo {volume} {18}},\ \bibinfo
  {pages} {7674} (\bibinfo {year} {2018})}\BibitemShut {NoStop}%
\bibitem [{\citenamefont {Aharoni}(1998)}]{Aharoni1998}%
  \BibitemOpen
  \bibfield  {author} {\bibinfo {author} {\bibfnamefont {A.}~\bibnamefont
  {Aharoni}},\ }\href@noop {} {\bibfield  {journal} {\bibinfo  {journal}
  {Journal of Applied Physics}\ }\textbf {\bibinfo {volume} {83}},\ \bibinfo
  {pages} {3432} (\bibinfo {year} {1998})}\BibitemShut {NoStop}%
\bibitem [{\citenamefont {Roy}\ \emph {et~al.}(2017)\citenamefont {Roy},
  \citenamefont {Wilson},\ and\ \citenamefont {Firstenberg}}]{Roy2017}%
  \BibitemOpen
  \bibfield  {author} {\bibinfo {author} {\bibfnamefont {D.}~\bibnamefont
  {Roy}}, \bibinfo {author} {\bibfnamefont {C.}~\bibnamefont {Wilson}},\ and\
  \bibinfo {author} {\bibfnamefont {O.}~\bibnamefont {Firstenberg}},\
  }\href@noop {} {\bibfield  {journal} {\bibinfo  {journal} {Reviews of Modern
  Physics}\ }\textbf {\bibinfo {volume} {89}},\ \bibinfo {pages} {021001}
  (\bibinfo {year} {2017})}\BibitemShut {NoStop}%
\bibitem [{\citenamefont {Zhao}\ \emph {et~al.}(2016)\citenamefont {Zhao},
  \citenamefont {Song}, \citenamefont {Yang}, \citenamefont {Su}, \citenamefont
  {Yuan}, \citenamefont {Parkin}, \citenamefont {Shi},\ and\ \citenamefont
  {Han}}]{Zhao2016}%
  \BibitemOpen
  \bibfield  {author} {\bibinfo {author} {\bibfnamefont {Y.}~\bibnamefont
  {Zhao}}, \bibinfo {author} {\bibfnamefont {Q.}~\bibnamefont {Song}}, \bibinfo
  {author} {\bibfnamefont {S.-H.}\ \bibnamefont {Yang}}, \bibinfo {author}
  {\bibfnamefont {T.}~\bibnamefont {Su}}, \bibinfo {author} {\bibfnamefont
  {W.}~\bibnamefont {Yuan}}, \bibinfo {author} {\bibfnamefont {S.~S.~P.}\
  \bibnamefont {Parkin}}, \bibinfo {author} {\bibfnamefont {J.}~\bibnamefont
  {Shi}},\ and\ \bibinfo {author} {\bibfnamefont {W.}~\bibnamefont {Han}},\
  }\href@noop {} {\bibfield  {journal} {\bibinfo  {journal} {Scientific
  Reports}\ }\textbf {\bibinfo {volume} {6}} (\bibinfo {year}
  {2016})}\BibitemShut {NoStop}%
\bibitem [{\citenamefont {G{\"o}ppl}\ \emph {et~al.}(2008)\citenamefont
  {G{\"o}ppl}, \citenamefont {Fragner}, \citenamefont {Baur}, \citenamefont
  {Bianchetti}, \citenamefont {Filipp}, \citenamefont {Fink}, \citenamefont
  {Leek}, \citenamefont {Puebla}, \citenamefont {Steffen},\ and\ \citenamefont
  {Wallraff}}]{Goeppl2008}%
  \BibitemOpen
  \bibfield  {author} {\bibinfo {author} {\bibfnamefont {M.}~\bibnamefont
  {G{\"o}ppl}}, \bibinfo {author} {\bibfnamefont {A.}~\bibnamefont {Fragner}},
  \bibinfo {author} {\bibfnamefont {M.}~\bibnamefont {Baur}}, \bibinfo {author}
  {\bibfnamefont {R.}~\bibnamefont {Bianchetti}}, \bibinfo {author}
  {\bibfnamefont {S.}~\bibnamefont {Filipp}}, \bibinfo {author} {\bibfnamefont
  {J.~M.}\ \bibnamefont {Fink}}, \bibinfo {author} {\bibfnamefont {P.~J.}\
  \bibnamefont {Leek}}, \bibinfo {author} {\bibfnamefont {G.}~\bibnamefont
  {Puebla}}, \bibinfo {author} {\bibfnamefont {L.}~\bibnamefont {Steffen}},\
  and\ \bibinfo {author} {\bibfnamefont {A.}~\bibnamefont {Wallraff}},\
  }\href@noop {} {\bibfield  {journal} {\bibinfo  {journal} {Journal of Applied
  Physics}\ }\textbf {\bibinfo {volume} {104}},\ \bibinfo {pages} {113904}
  (\bibinfo {year} {2008})}\BibitemShut {NoStop}%
\bibitem [{\citenamefont {Fano}(1961)}]{Fano1961}%
  \BibitemOpen
  \bibfield  {author} {\bibinfo {author} {\bibfnamefont {U.}~\bibnamefont
  {Fano}},\ }\href@noop {} {\bibfield  {journal} {\bibinfo  {journal} {Physical
  Review}\ }\textbf {\bibinfo {volume} {124}},\ \bibinfo {pages} {1866}
  (\bibinfo {year} {1961})}\BibitemShut {NoStop}%
\bibitem [{\citenamefont {Bushev}\ \emph {et~al.}(2011)\citenamefont {Bushev},
  \citenamefont {Feofanov}, \citenamefont {Rotzinger}, \citenamefont
  {Protopopov}, \citenamefont {Cole}, \citenamefont {Wilson}, \citenamefont
  {Fischer}, \citenamefont {Lukashenko},\ and\ \citenamefont
  {Ustinov}}]{Bushev2011}%
  \BibitemOpen
  \bibfield  {author} {\bibinfo {author} {\bibfnamefont {P.}~\bibnamefont
  {Bushev}}, \bibinfo {author} {\bibfnamefont {A.~K.}\ \bibnamefont
  {Feofanov}}, \bibinfo {author} {\bibfnamefont {H.}~\bibnamefont {Rotzinger}},
  \bibinfo {author} {\bibfnamefont {I.}~\bibnamefont {Protopopov}}, \bibinfo
  {author} {\bibfnamefont {J.~H.}\ \bibnamefont {Cole}}, \bibinfo {author}
  {\bibfnamefont {C.~M.}\ \bibnamefont {Wilson}}, \bibinfo {author}
  {\bibfnamefont {G.}~\bibnamefont {Fischer}}, \bibinfo {author} {\bibfnamefont
  {A.}~\bibnamefont {Lukashenko}},\ and\ \bibinfo {author} {\bibfnamefont
  {A.~V.}\ \bibnamefont {Ustinov}},\ }\href@noop {} {\bibfield  {journal}
  {\bibinfo  {journal} {Physical Review B}\ }\textbf {\bibinfo {volume} {84}},\
  \bibinfo {pages} {060501} (\bibinfo {year} {2011})}\BibitemShut {NoStop}%
\bibitem [{\citenamefont {Mart{\'{\i}}nez-P{\'{e}}rez}\ and\ \citenamefont
  {Zueco}(2019)}]{MartinezPerez2019}%
  \BibitemOpen
  \bibfield  {author} {\bibinfo {author} {\bibfnamefont {M.~J.}\ \bibnamefont
  {Mart{\'{\i}}nez-P{\'{e}}rez}}\ and\ \bibinfo {author} {\bibfnamefont
  {D.}~\bibnamefont {Zueco}},\ }\href@noop {} {\bibfield  {journal} {\bibinfo
  {journal} {New Journal of Physics}\ }\textbf {\bibinfo {volume} {21}},\
  \bibinfo {pages} {115002} (\bibinfo {year} {2019})}\BibitemShut {NoStop}%
\bibitem [{\citenamefont {Mart{\'{\i}}nez-P{\'{e}}rez}\ and\ \citenamefont
  {Zueco}(2018)}]{MartinezPerez2018}%
  \BibitemOpen
  \bibfield  {author} {\bibinfo {author} {\bibfnamefont {M.~J.}\ \bibnamefont
  {Mart{\'{\i}}nez-P{\'{e}}rez}}\ and\ \bibinfo {author} {\bibfnamefont
  {D.}~\bibnamefont {Zueco}},\ }\href@noop {} {\bibfield  {journal} {\bibinfo
  {journal} {{ACS} Photonics}\ }\textbf {\bibinfo {volume} {6}},\ \bibinfo
  {pages} {360} (\bibinfo {year} {2018})}\BibitemShut {NoStop}%
\bibitem [{\citenamefont {H{\"u}mmer}\ \emph {et~al.}(2012)\citenamefont
  {H{\"u}mmer}, \citenamefont {Reuther}, \citenamefont {H{\"a}nggi},\ and\
  \citenamefont {Zueco}}]{Huemmer2012}%
  \BibitemOpen
  \bibfield  {author} {\bibinfo {author} {\bibfnamefont {T.}~\bibnamefont
  {H{\"u}mmer}}, \bibinfo {author} {\bibfnamefont {G.~M.}\ \bibnamefont
  {Reuther}}, \bibinfo {author} {\bibfnamefont {P.}~\bibnamefont
  {H{\"a}nggi}},\ and\ \bibinfo {author} {\bibfnamefont {D.}~\bibnamefont
  {Zueco}},\ }\href@noop {} {\bibfield  {journal} {\bibinfo  {journal}
  {Physical Review A}\ }\textbf {\bibinfo {volume} {85}},\ \bibinfo {pages}
  {052320} (\bibinfo {year} {2012})}\BibitemShut {NoStop}%
\bibitem [{\citenamefont {Mac{\^{e}}do}\ \emph {et~al.}(2021)\citenamefont
  {Mac{\^{e}}do}, \citenamefont {Holland}, \citenamefont {Baity}, \citenamefont
  {McLellan}, \citenamefont {Livesey}, \citenamefont {Stamps}, \citenamefont
  {Weides},\ and\ \citenamefont {Bozhko}}]{Macedo2021}%
  \BibitemOpen
  \bibfield  {author} {\bibinfo {author} {\bibfnamefont {R.}~\bibnamefont
  {Mac{\^{e}}do}}, \bibinfo {author} {\bibfnamefont {R.~C.}\ \bibnamefont
  {Holland}}, \bibinfo {author} {\bibfnamefont {P.~G.}\ \bibnamefont {Baity}},
  \bibinfo {author} {\bibfnamefont {L.~J.}\ \bibnamefont {McLellan}}, \bibinfo
  {author} {\bibfnamefont {K.~L.}\ \bibnamefont {Livesey}}, \bibinfo {author}
  {\bibfnamefont {R.~L.}\ \bibnamefont {Stamps}}, \bibinfo {author}
  {\bibfnamefont {M.~P.}\ \bibnamefont {Weides}},\ and\ \bibinfo {author}
  {\bibfnamefont {D.~A.}\ \bibnamefont {Bozhko}},\ }\href@noop {} {\bibfield
  {journal} {\bibinfo  {journal} {Physical Review Applied}\ }\textbf {\bibinfo
  {volume} {15}},\ \bibinfo {pages} {024065} (\bibinfo {year}
  {2021})}\BibitemShut {NoStop}%
\bibitem [{\citenamefont {Bourhill}\ \emph {et~al.}(2020)\citenamefont
  {Bourhill}, \citenamefont {Castel}, \citenamefont {Manchec},\ and\
  \citenamefont {Cochet}}]{Bourhill2020}%
  \BibitemOpen
  \bibfield  {author} {\bibinfo {author} {\bibfnamefont {J.}~\bibnamefont
  {Bourhill}}, \bibinfo {author} {\bibfnamefont {V.}~\bibnamefont {Castel}},
  \bibinfo {author} {\bibfnamefont {A.}~\bibnamefont {Manchec}},\ and\ \bibinfo
  {author} {\bibfnamefont {G.}~\bibnamefont {Cochet}},\ }\href@noop {}
  {\bibfield  {journal} {\bibinfo  {journal} {Journal of Applied Physics}\
  }\textbf {\bibinfo {volume} {128}},\ \bibinfo {pages} {073904} (\bibinfo
  {year} {2020})}\BibitemShut {NoStop}%
\bibitem [{\citenamefont {Khapaev}\ \emph {et~al.}(2002)\citenamefont
  {Khapaev}, \citenamefont {Kupriyanov}, \citenamefont {Goldobin},\ and\
  \citenamefont {Siegel}}]{Khapaev2002}%
  \BibitemOpen
  \bibfield  {author} {\bibinfo {author} {\bibfnamefont {M.~M.}\ \bibnamefont
  {Khapaev}}, \bibinfo {author} {\bibfnamefont {M.~Y.}\ \bibnamefont
  {Kupriyanov}}, \bibinfo {author} {\bibfnamefont {E.}~\bibnamefont
  {Goldobin}},\ and\ \bibinfo {author} {\bibfnamefont {M.}~\bibnamefont
  {Siegel}},\ }\href@noop {} {\bibfield  {journal} {\bibinfo  {journal}
  {Superconductor Science and Technology}\ }\textbf {\bibinfo {volume} {16}},\
  \bibinfo {pages} {24} (\bibinfo {year} {2002})}\BibitemShut {NoStop}%
\bibitem [{\citenamefont {Jenkins}\ \emph {et~al.}(2013)\citenamefont
  {Jenkins}, \citenamefont {HÃ¼mmer}, \citenamefont
  {Mart{\'{\i}}nez-P{\'{e}}rez}, \citenamefont {Garc{\'{\i}}a-Ripoll},
  \citenamefont {Zueco},\ and\ \citenamefont {Luis}}]{Jenkins2013}%
  \BibitemOpen
  \bibfield  {author} {\bibinfo {author} {\bibfnamefont {M.}~\bibnamefont
  {Jenkins}}, \bibinfo {author} {\bibfnamefont {T.}~\bibnamefont {HÃ¼mmer}},
  \bibinfo {author} {\bibfnamefont {M.~J.}\ \bibnamefont
  {Mart{\'{\i}}nez-P{\'{e}}rez}}, \bibinfo {author} {\bibfnamefont
  {J.}~\bibnamefont {Garc{\'{\i}}a-Ripoll}}, \bibinfo {author} {\bibfnamefont
  {D.}~\bibnamefont {Zueco}},\ and\ \bibinfo {author} {\bibfnamefont
  {F.}~\bibnamefont {Luis}},\ }\href@noop {} {\bibfield  {journal} {\bibinfo
  {journal} {New Journal of Physics}\ }\textbf {\bibinfo {volume} {15}},\
  \bibinfo {pages} {095007} (\bibinfo {year} {2013})}\BibitemShut {NoStop}%
\bibitem [{\citenamefont {Vansteenkiste}\ \emph {et~al.}(2014)\citenamefont
  {Vansteenkiste}, \citenamefont {Leliaert}, \citenamefont {Dvornik},
  \citenamefont {Helsen}, \citenamefont {Garcia-Sanchez},\ and\ \citenamefont
  {Waeyenberge}}]{Vansteenkiste2014}%
  \BibitemOpen
  \bibfield  {author} {\bibinfo {author} {\bibfnamefont {A.}~\bibnamefont
  {Vansteenkiste}}, \bibinfo {author} {\bibfnamefont {J.}~\bibnamefont
  {Leliaert}}, \bibinfo {author} {\bibfnamefont {M.}~\bibnamefont {Dvornik}},
  \bibinfo {author} {\bibfnamefont {M.}~\bibnamefont {Helsen}}, \bibinfo
  {author} {\bibfnamefont {F.}~\bibnamefont {Garcia-Sanchez}},\ and\ \bibinfo
  {author} {\bibfnamefont {B.~V.}\ \bibnamefont {Waeyenberge}},\ }\href@noop {}
  {\bibfield  {journal} {\bibinfo  {journal} {{AIP} Advances}\ }\textbf
  {\bibinfo {volume} {4}},\ \bibinfo {pages} {107133} (\bibinfo {year}
  {2014})}\BibitemShut {NoStop}%
\end{thebibliography}%

\end{document}